\documentclass[aps,prl,twocolumn,superscriptaddress,longbibliography]{revtex4-2}

\usepackage[dvipdfmx]{graphicx}
\usepackage[small,bf]{caption}
\usepackage[subrefformat=parens, labelformat=parens]{subcaption}
\usepackage{color}
\usepackage{physics}
\usepackage[version=3]{mhchem}
\usepackage{lineno}

\begin{document}

\title{Massive-mode polarization entangled biphoton frequency comb}

\author{Tomohiro Yamazaki}
\affiliation{Graduate School of Engineering Science, Osaka University, Toyonaka, Osaka 560-8531, Japan}

\author{Rikizo Ikuta}
\email{ikuta@mp.es.osaka-u.ac.jp}
\affiliation{Graduate School of Engineering Science, Osaka University, Toyonaka, Osaka 560-8531, Japan}
\affiliation{Center for Quantum Information and Quantum Biology, Osaka University, Osaka 560-8531, Japan}

\author{Toshiki Kobayashi}
\affiliation{Graduate School of Engineering Science, Osaka University, Toyonaka, Osaka 560-8531, Japan}
\affiliation{Center for Quantum Information and Quantum Biology, Osaka University, Osaka 560-8531, Japan}

\author{Shigehito Miki}
\affiliation{Advanced ICT Research Institute, National Institute of Information and Communications Technology (NICT), Kobe 651-2492, Japan}
\affiliation{Graduate School of Engineering, Kobe University, Kobe 657-0013, Japan}

\author{Fumihiro China}
\affiliation{Advanced ICT Research Institute, National Institute of Information and Communications Technology (NICT), Kobe 651-2492, Japan}

\author{Hirotaka Terai}
\affiliation{Advanced ICT Research Institute, National Institute of Information and Communications Technology (NICT), Kobe 651-2492, Japan}

\author{Nobuyuki Imoto}
\affiliation{Center for Quantum Information and Quantum Biology, Osaka University, Osaka 560-8531, Japan}

\author{Takashi Yamamoto}
\affiliation{Graduate School of Engineering Science, Osaka University, Toyonaka, Osaka 560-8531, Japan}
\affiliation{Center for Quantum Information and Quantum Biology, Osaka University, Osaka 560-8531, Japan} 

\date{\today} 

\begin{abstract}
    A frequency-multiplexed entangled photon pair and a high-dimensional hyperentangled photon pair are useful to realize a high-capacity quantum communication.
    A biphoton frequency comb (BFC) with entanglement can be used to prepare both states.
    We demonstrate polarization entangled BFCs with over 1400 frequency modes, which is approximately two orders of magnitude larger than those of earlier entangled BFCs, by placing a singly resonant periodically poled \ce{LiNbO3} waveguide resonator within a Sagnac loop.
    The BFCs are demonstrated by measuring the joint spectral intensity, cross-correlation, and autocorrelation. Moreover, the polarization entanglement at representative groups of frequency modes is verified by quantum state tomography, where each fidelity is over 0.7.
    The efficient generation of a massive-mode entangled BFC is expected to accelerate the increase of capacity in quantum communication.
\end{abstract}


\maketitle

\section{Introduction} 
Entangled photon pairs are essential resources for quantum information processing such as quantum communication~\cite{Gisin2007,Yin2017},
quantum computation~\cite{Kok2007,Lukens2016a}, and quantum metrology~\cite{Giovannetti2011}. Fully utilizing the degrees of freedom (DOFs) of photons remain a crucial challenge in developing better photon resources.
Recently, the quantum nature of the frequency DOF of photons has been studied intensively~\cite{Kobayashi2016a,Kues2017}.
To this end, a quantum frequency comb~\cite{Caspani2016b,Kues2019}, which is the superposition of different frequency modes at a certain mode spacing, 
is gaining research interest owing to its advantage in the generation~\cite{Imany2018a,Ikuta2019d,Maltese2020} and manipulation~\cite{Lu2018b,Lu2019b,Reimer2019,Ikuta2021} of high-dimensional quantum states. A quantum frequency comb formed by a frequency-entangled photon pair, i.e. a biphoton frequency comb (BFC), is a building block of quantum information processing that relies on its frequency DOF.

BFCs are generated efficiently via spontaneous parametric down-conversion (SPDC) or spontaneous four-wave mixing (SFWM) inside a cavity. 
On the one hand, a photon pair generated by SPDC/SFWM without a cavity has a broad bandwidth, which can be determined by the phase-matching condition{.
On the other hand, photon pair generation in SPDC/SFWM} inside a cavity occurs intensively in the resonant frequency modes of the cavity; this results in the highly efficient generation of BFCs. Thus far, BFCs have mostly been generated using SFWM~\cite{Reimer2014,Reimer2016,Kues2017,Fujiwara2017,Imany2018a} and SPDC~\cite{Fortsch2013c, Guo2017} with micro-ring resonators~\cite{Caspani2017}.
However, the number of their frequency modes is limited to {a few dozen} because of the smallness of the cavity size and ``the cluster effect", which is the suppression of photon pair generation in a certain frequency range, in a doubly resonant configuration~\cite{Pomarico2009a,Pomarico2012}. In our experiment, we use a monolithic periodically poled \ce{LiNbO3} integrated waveguide resonator (PPLN/WR). Its cavity confines only the longer-wavelength (signal) photons of the photon pair, and it does not confine shorter-wavelength (idler) photons to avoid the cluster effect{~\cite{Ikuta2019d}}; thus, a BFC with over 1400 frequency modes is achieved.
{We demonstrate the generation of polarization entangled BFC ranging over 1400 modes, which is about two orders of magnitude larger than previous entangled BFCs, by placing the singly resonant PPLN/WR into a Sagnac interferometer. In the Sagnac interferometer, the BFCs for horizontal and vertical polarizations are generated by the same PPLN/WR, and therefore, a perfect spectral overlap between the H- and V-polarized BFCs can be ensured.}

\nolinenumbers
\begin{figure*}[tbp]
    \centering
    \includegraphics[scale=0.21, trim=0 150 0 100, clip]{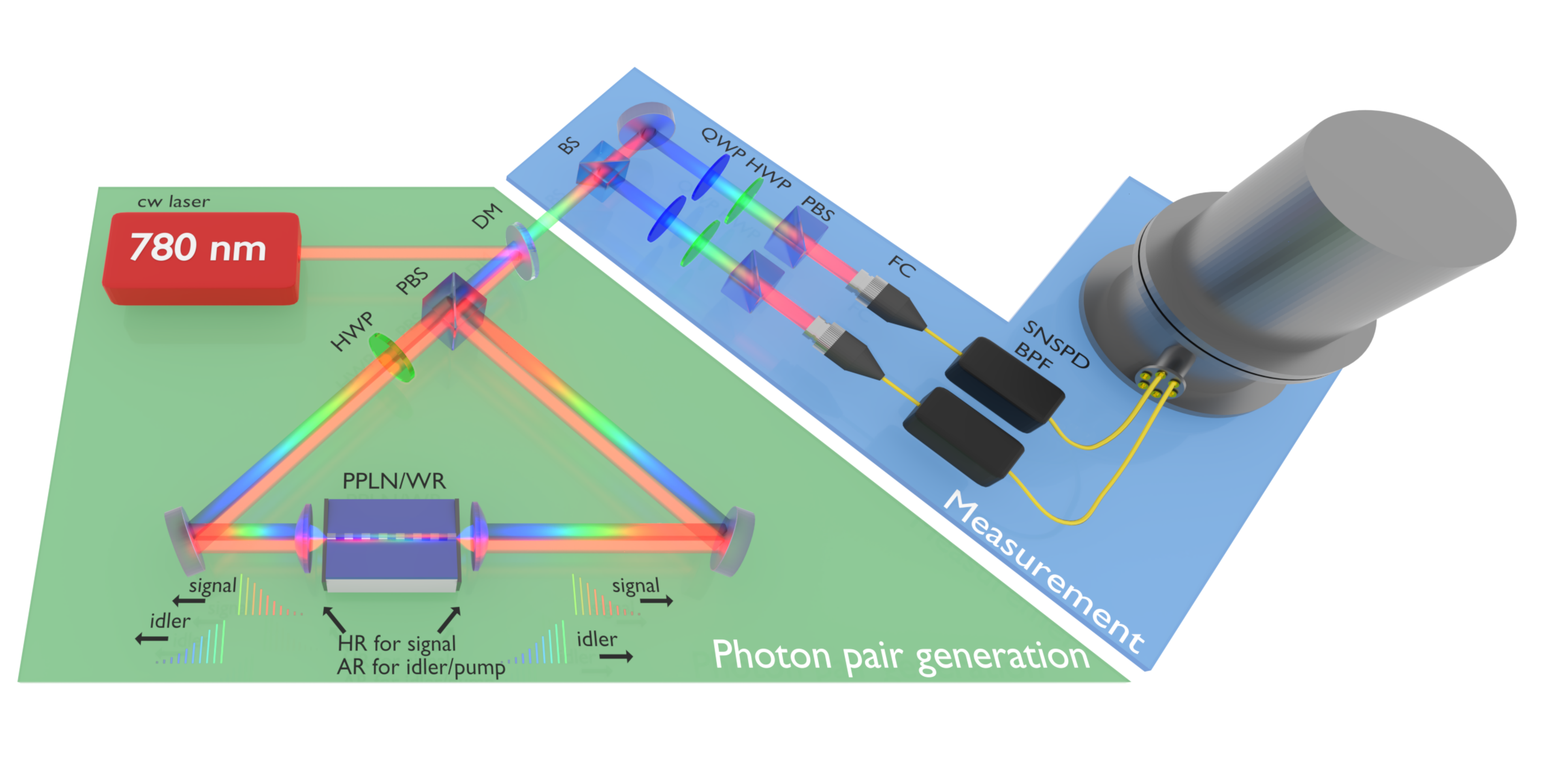}
    \caption{{\bf Experimental setup for the generation and measurement of quantum states.} BS: beam splitter; PBS: polarizing beam splitter; HWP: half-wave plate; QWP: quarter-wave plate; DM: dichroic mirror; FC: fiber coupler; BPF: bandpass filter; SNSPD: superconducting nanowire single-photon detector {; PPLN/WR, periodically poled \ce{LiNbO3} waveguide resonator.}}
    \label{setup}
\end{figure*}

Additional entanglement in another DOF of the BFC extends the use case to various applications such as (a) high-dimensional hyperentangled photon pairs~\cite{Xie2015,Reimer2019,Chang:2021vm} or (b) frequency-multiplexed entangled photon pairs~\cite{Reimer2016}.
{Suppose} a simplified model for generating polarization entangled BFCs based on SPDC, where the generated state is written by 
\begin{equation}
    \ket{\Psi} = \left(\prod_{m=1}^M S_{m,H}(\zeta) S_{m,V}(\zeta)\right) \ket{0}.
    \label{PBFC}
\end{equation}
$S_{m,H(V)}(\zeta)$ represents a two-mode squeezing operator where $m~(1\leq m\leq M)$ 
and H(V) denote the frequency mode index and the horizontal (vertical) polarization, respectively. The complex squeezing parameter $\zeta$ is proportional to the complex amplitude of the SPDC pump light.

From equation~\eqref{PBFC},
we prepare two different states (a) and (b)
by adjusting only the pump power through parameter $\zeta$ as follows:
(i) In the low pump power regime satisfying $2M \abs{\zeta}^2 \ll 1$, 
a single photon pair state is represented by
\begin{equation}
    \ket{\Psi} \simeq \frac{1}{\sqrt{2}} \qty(\ket{HH}_{i,s}+\ket{VV}_{i,s})\otimes \frac{1}{\sqrt{M}} \qty(\sum_{m=1}^M \ket{\omega_{m} \omega_{m}}_{i,s}),
    \label{HHEP}
\end{equation}
where $\ket{\omega_{i(s),m}}$ represents the state of the idler (signal) photon in the frequency index $m$.
This state corresponds to a high-dimensional hyperentangled state (a) of polarization and frequency, which is valuable for many protocols with hyperentanglement~\cite{Deng2017,Kwiat1998,Barreiro2008,Sheng2010}.
(ii) For $2 \abs{\zeta}^2 \ll 1$,
the generated state in each frequency mode is approximately a single photon pair, and it is represented as
\begin{equation}
    \ket{\Psi} \propto \bigotimes_{m=1}^M \qty[\ket{0}_m+\zeta\qty(\ket{HH}_{{i,m},{s,m}}+\ket{VV}_{{i,m},{s,m}})].
    \label{FMEP}
\end{equation}
This state is equivalent to the state formed by many polarization-entangled photon pairs with different frequency indices that share a single pump light. This frequency-multiplexed entangled photon pair (b) can be applied to broadband entanglement-distribution network~\cite{Lim2010a, Wengerowsky2018, Joshi2020-jz, Lingaraju2021-iv, Appas2021-nc, Alshowkan2021-ut}.
{Conditions of the pump power of (a) and (b) are different by $M$, which becomes remarkable in a massive mode BFC. In our experiment, we adjust the pump power based on the frequency window of the measurements. A realistic model for polarization-entangled BFCs is discussed in the Supplementary Information.} 
    
\nolinenumbers
\begin{figure*}[tbp]
    \begin{tabular}{lr}
        \begin{minipage}[c]{0.60\linewidth}
            \centering
            \begin{tabular}{cc}
                \raisebox{-\height}{\bf a} & \raisebox{-\height}{\includegraphics[scale=0.70, trim=0 0 0 40, clip]{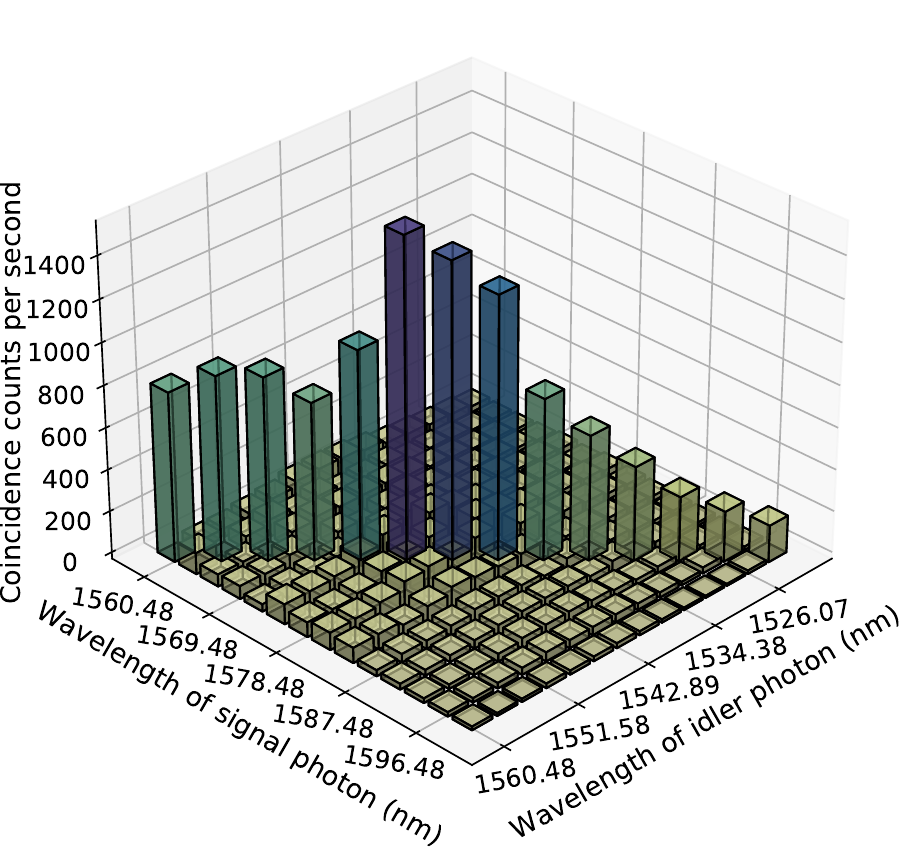}}
            \end{tabular}
        \end{minipage}

        \begin{minipage}[c]{0.40\linewidth}
            \centering
            \begin{tabular}{cc}
                \raisebox{-\height}{\bf b} & \raisebox{-\height}{\includegraphics[scale=0.4, trim=0 0 0 60, clip]{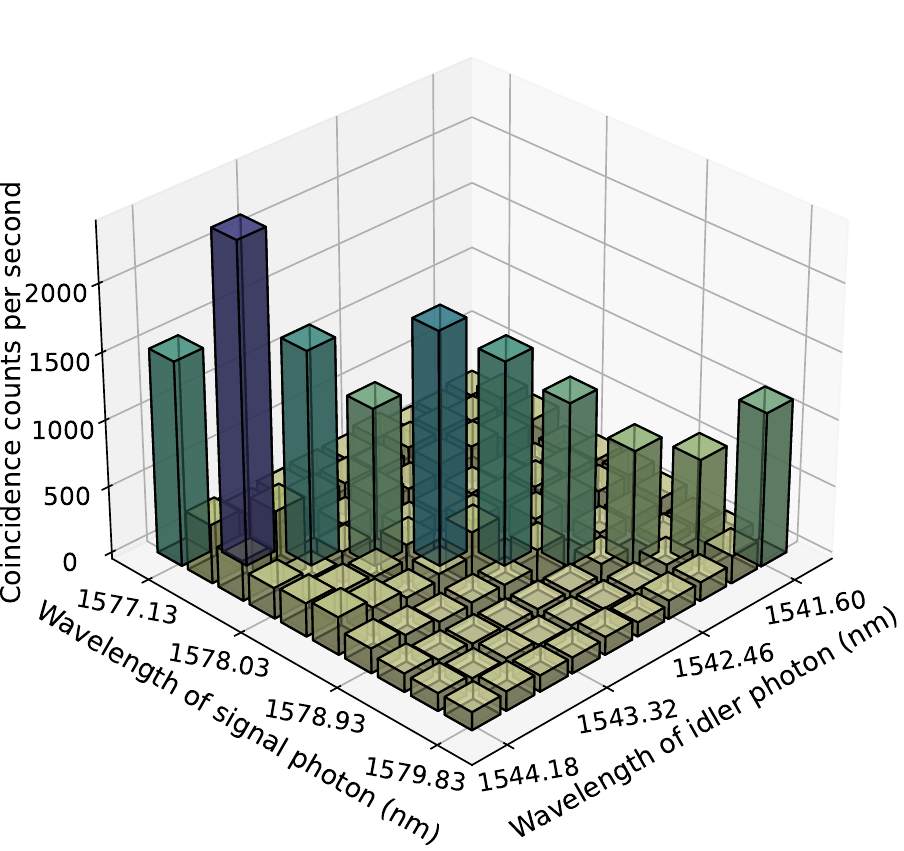}} \\
                \raisebox{-\height}{\bf c} & \raisebox{-\height}{\includegraphics[scale=0.4, trim=0 0 0 60, clip]{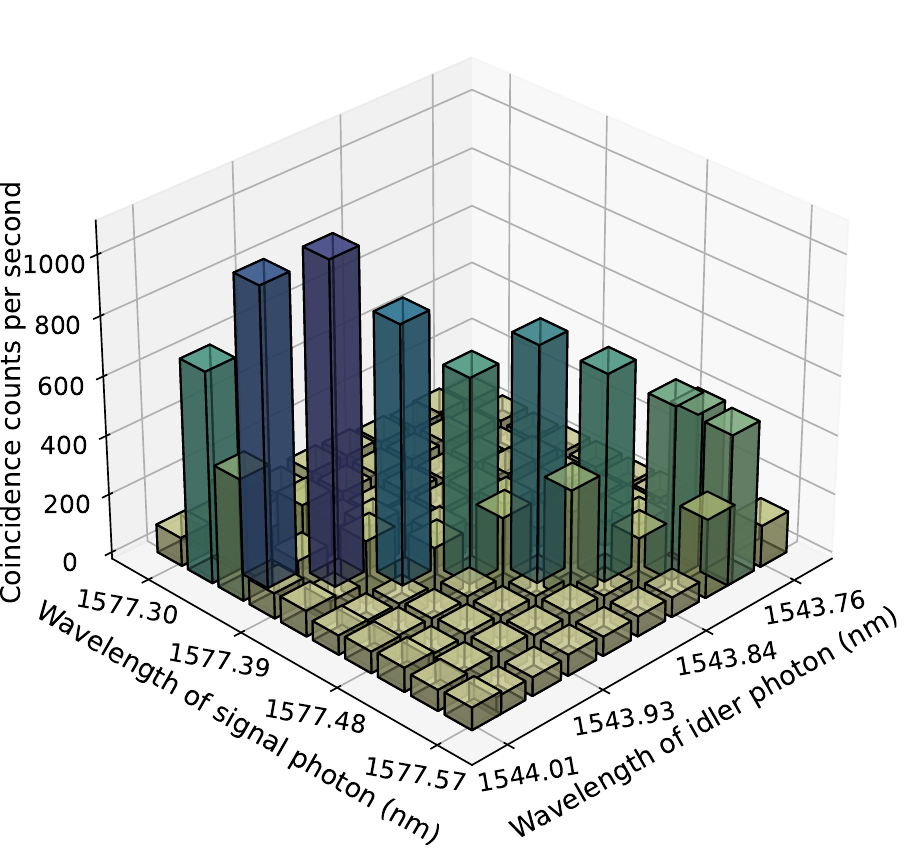}}
            \end{tabular}
        \end{minipage}
    \end{tabular}
    \caption{{\bf Joint spectral intensity.} 
    Coincidence counts among frequency modes with spectral resolutions of {\bf a} 3.00 nm, {\bf b} 0.30 nm, and {\bf c} 0.03 nm are shown. The resolutions are selected using the BPFs. The pump powers are adjusted based on the resolution to 50~$\mu$W, 500~$\mu$W, and 5~mW for {\bf a}, {\bf b}, and {\bf c}, respectively.
    Peaks appear at the specific frequency combinations that satisfy the energy conservation with the pump, signal, and idler photons in all frequency ranges measured: {\bf a} [1560.48 nm, 1599.48 nm] and [1522.43 nm, 1560.48 nm], {\bf b} [1577.13 nm, 1579.83 nm] and [1541.60 nm, 1544.18 nm], and {\bf c} [1577.30 nm, 1577.57 nm] and [1543.76 nm, 1544.01 nm]. The range of {\bf b} is equivalent to the spectral resolution of {\bf a} and the range of {\bf c} is the spectral resolution of {\bf b}.}
    \label{Z}
\end{figure*}

\section{Results} 
Figure 1 shows the experimental setup. We used a continuous-wave (CW) pump laser at 780.24~nm, whose power is adjustable from 50~$\mu$W to 5~mW. The Sagnac interferometer comprises a polarizing beam splitter (PBS), PPLN/WR, and half-wave plate (HWP). In the Sagnac interferometer, $\ket{HH}$ and $\ket{VV}$ photon pairs are generated in the clockwise and anticlockwise directions, respectively. The photon pairs are mixed by PBS and propagated in the direction opposite to that of the pump laser. The generated quantum state is a polarization-entangled BFC, which corresponds to equation~\eqref{PBFC}. {The setup does not require any stabilization besides the temperature control of the PPLN/WR because of the monolithic structure of PPLN/WR and the phase stability of a Sagnac interferometer~\cite{Shi2004,Kim2006}.}

The two photons are divided using a 50:50 beam splitter (BS) to evaluate the quantum states. Each photon goes to the setup for polarization quantum state tomography with a tunable bandpass filter (BPF). Then, they are detected using superconducting nanowire single-photon detectors (SNSPDs)~\cite{Miki2017}.
In the Sagnac interferometer, the undesired photon pairs $\ket{HV}$ and $\ket{VH}$ are generated by the signal components reflected at the resonator. We postselect only the desired photon pairs composed of the idler photons and the transmitted signal photons by time-resolved coincidence measurements by placing the PPLN/WR off-center from the interferometer (see Supplementary Information for details).

To observe the frequency-correlated photon pair generation over a wide spectral range, we first measure the joint spectral intensity (JSI) over 80~nm (1520-- 1600~nm) with a resolution of 3.00~nm, as shown in Fig.~2a. The JSIs with higher resolutions (0.30~nm and 0.03~nm) are shown in Figs.~2b and 2c, respectively. The peaks of the coincidence counts clearly obey the energy conservation and exhibit a frequency correlation in all frequency ranges.
The non-uniform shape of the peaks is qualitatively explained by the slightly remaining cluster effect attributed to the imperfection of the singly resonant configuration around the degenerate point of 1560.48~nm, as indicated in Fig.~S1f in the Supplementary Information.


The temporal second-order cross-correlation between the signal and idler photons illustrates an additional spectral property of the quantum frequency comb. Figure 3 shows an example of the results for the central wavelengths of 1580.48~nm and 1540.98~nm (see Supplementary Information for other frequency ranges).
The cross-correlation with a 3.00~nm filter bandwidth in Fig.~3a shows a beat signal reflecting the superposition of multiple frequency modes. A beat frequency of 3.5~GHz corresponds to the free spectral range (FSR) of the cavity. The beat signal in the delayed coincidence counts disappears when a single frequency mode is picked up by the filters with a 0.03~nm bandwidth corresponding to 3.5--3.7~GHz, as shown in Fig.~3b. We estimate the full width at half maximum (FWHM) of the linewidth of the frequency mode at approximately 1580~nm to be $\Delta f_{1/2}=$126~MHz based on the decay time of the cross-correlation. This implies that the comb teeth are well separated from each other because of the strong confinement of the signal photons in the resonator. 
In addition to the results shown in Fig.~2, wherein the photon pairs appear over 80~nm (1520--1600~nm), the clear beat signal implies the existence of frequency entanglement in the BFC with a frequency mode number of $M\sim 1400$.

\nolinenumbers

\begin{figure*}[t]
    \centering
    \begin{tabular}{cccc}
        \raisebox{-\height}{\bf a} & \raisebox{-\height}{\includegraphics[scale=0.35, trim=0 0 0 0, clip]{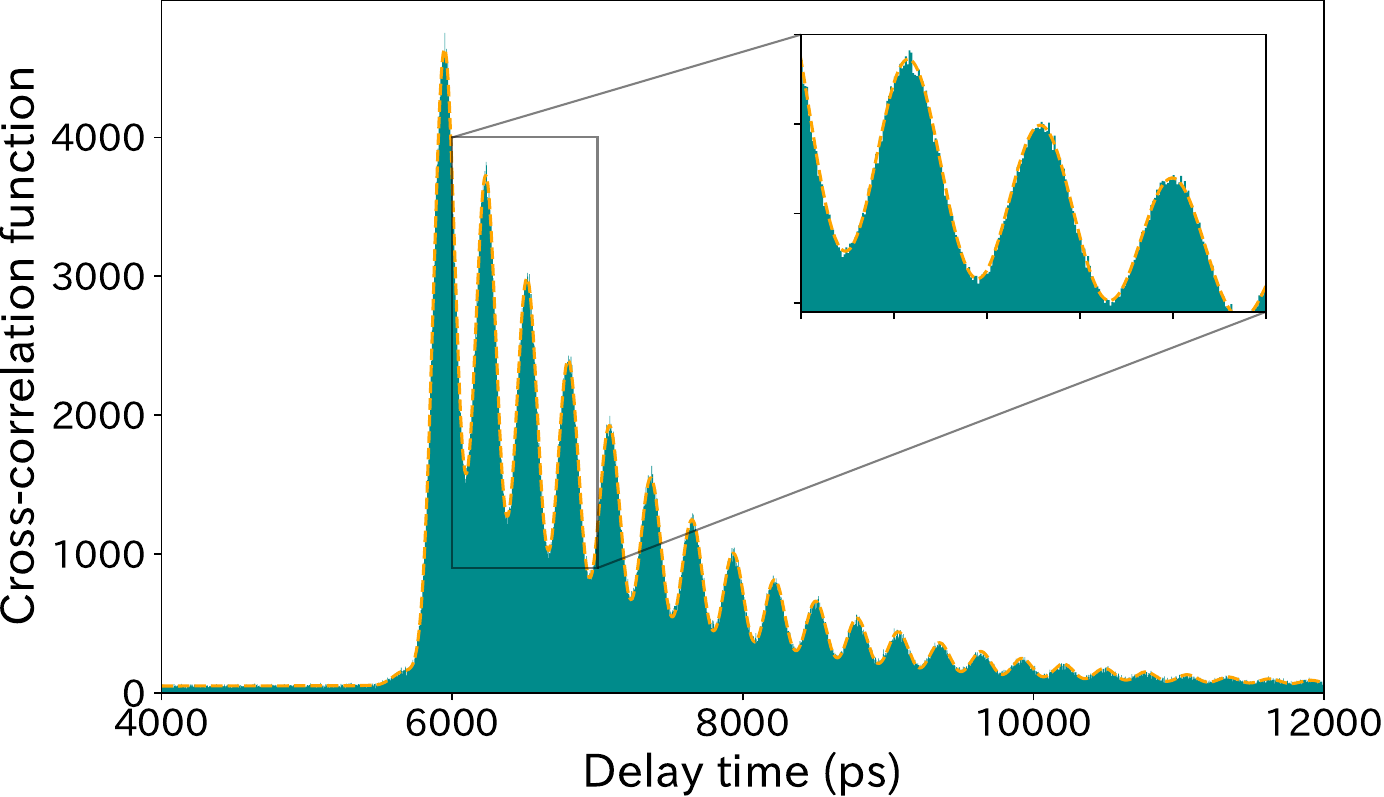}}
        \raisebox{-\height}{\bf b} & \raisebox{-\height}{\includegraphics[scale=0.35, trim=0 0 0 0, clip]{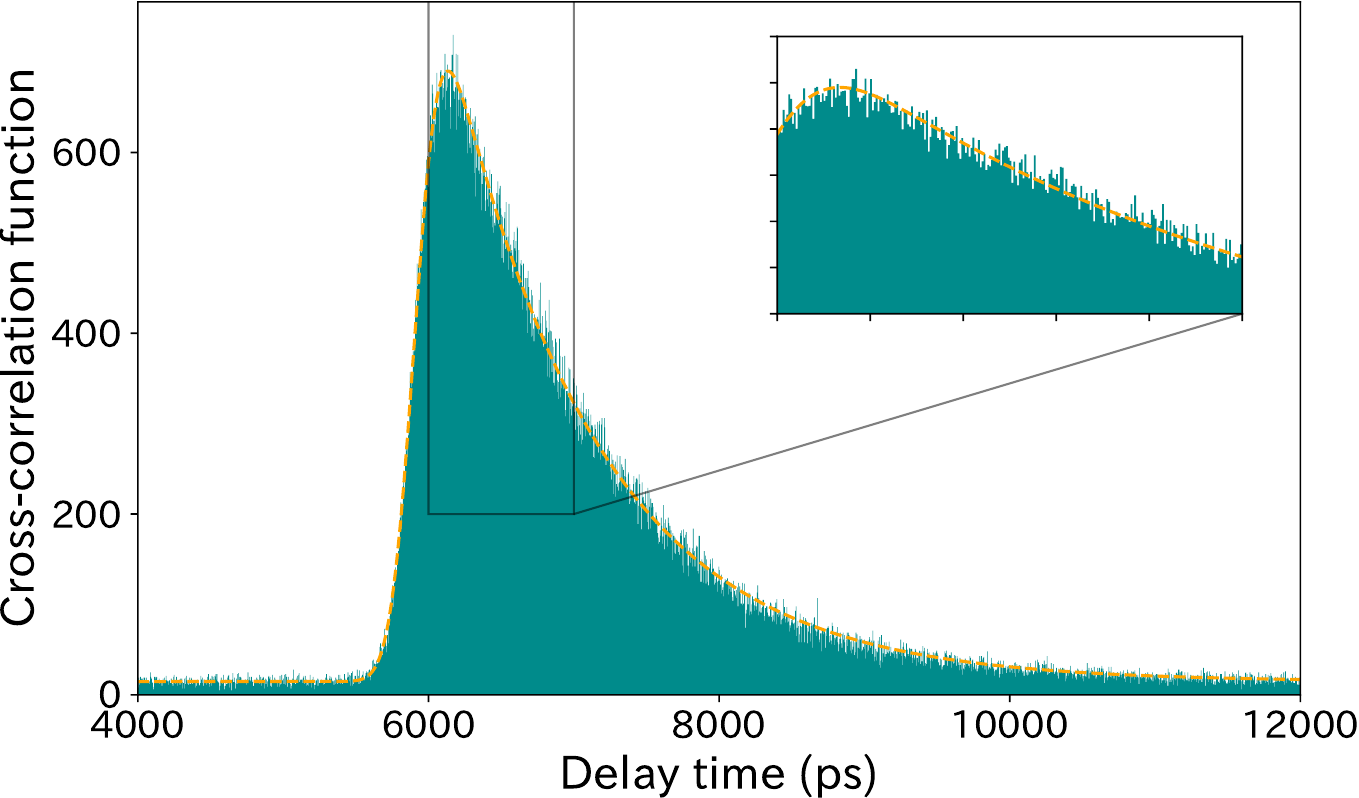}} &
    \end{tabular}
    \caption{{\bf Temporal second-order cross-correlation.} The delayed coincidence counts between the signal and idler photons with 4-ps time resolution, which correspond to the temporal second-order cross-correlation function. 
    The polarizations are fixed in the horizontal direction, and the wavelengths of the signal and idler photons are 1580.48~nm and 1540.98~nm, respectively. 
    The bandwidths of the BPFs are {\bf a} 3.00~nm ($\sim$ 100 frequency mode) and {\bf b} 0.03~nm (single-frequency mode).
    The data recording time is 1000~s.
    In {\bf a}, we observe the beating of coincidence counts, which is significant evidence of the quantum frequency comb. }
    \label{htm}
\end{figure*}

\begin{figure*}[tbp]
    \centering
    \begin{tabular}{cccc}
        \raisebox{-\height} {\bf a} & \raisebox{-\height}{\includegraphics[scale=0.35, trim=0 0 0 0, clip]{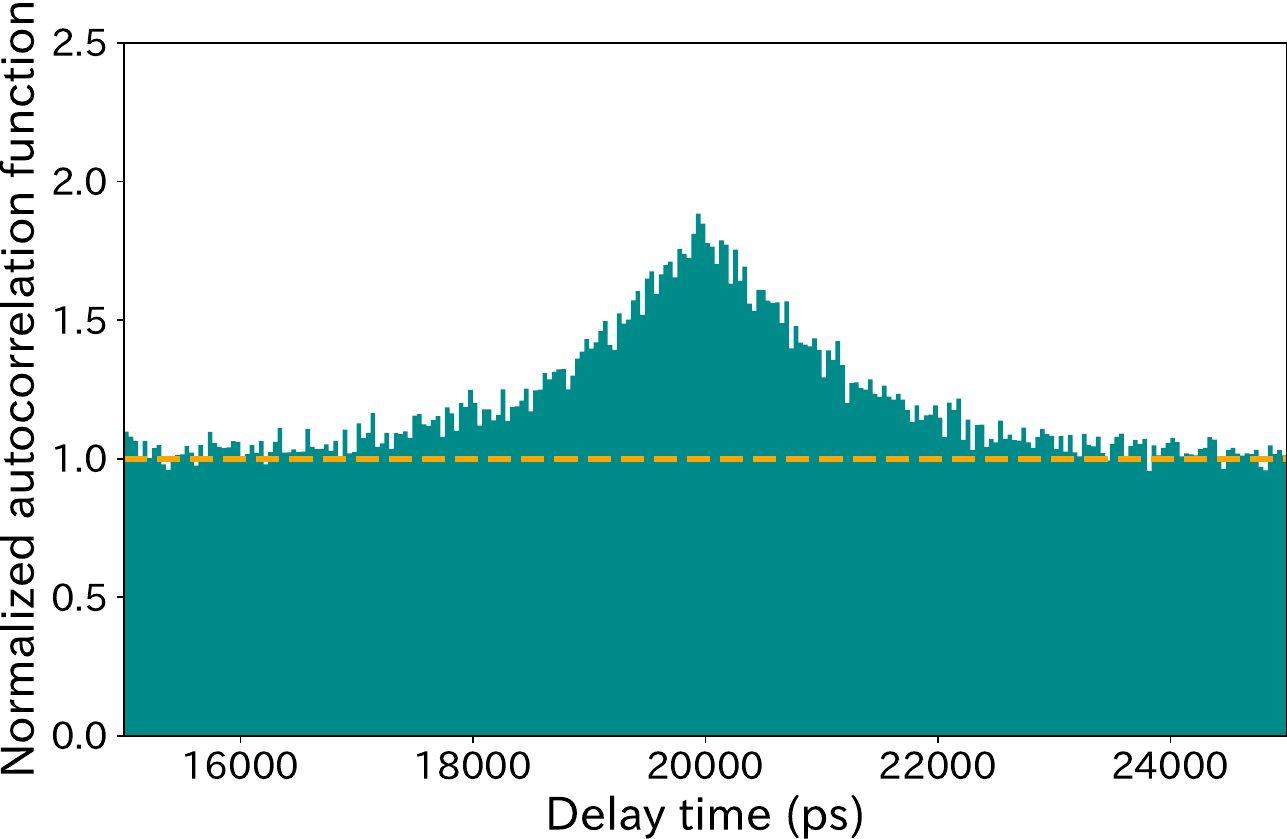}} &
        \raisebox{-\height} {\bf b} & \raisebox{-\height}{\includegraphics[scale=0.35, trim=0 0 0 0, clip]{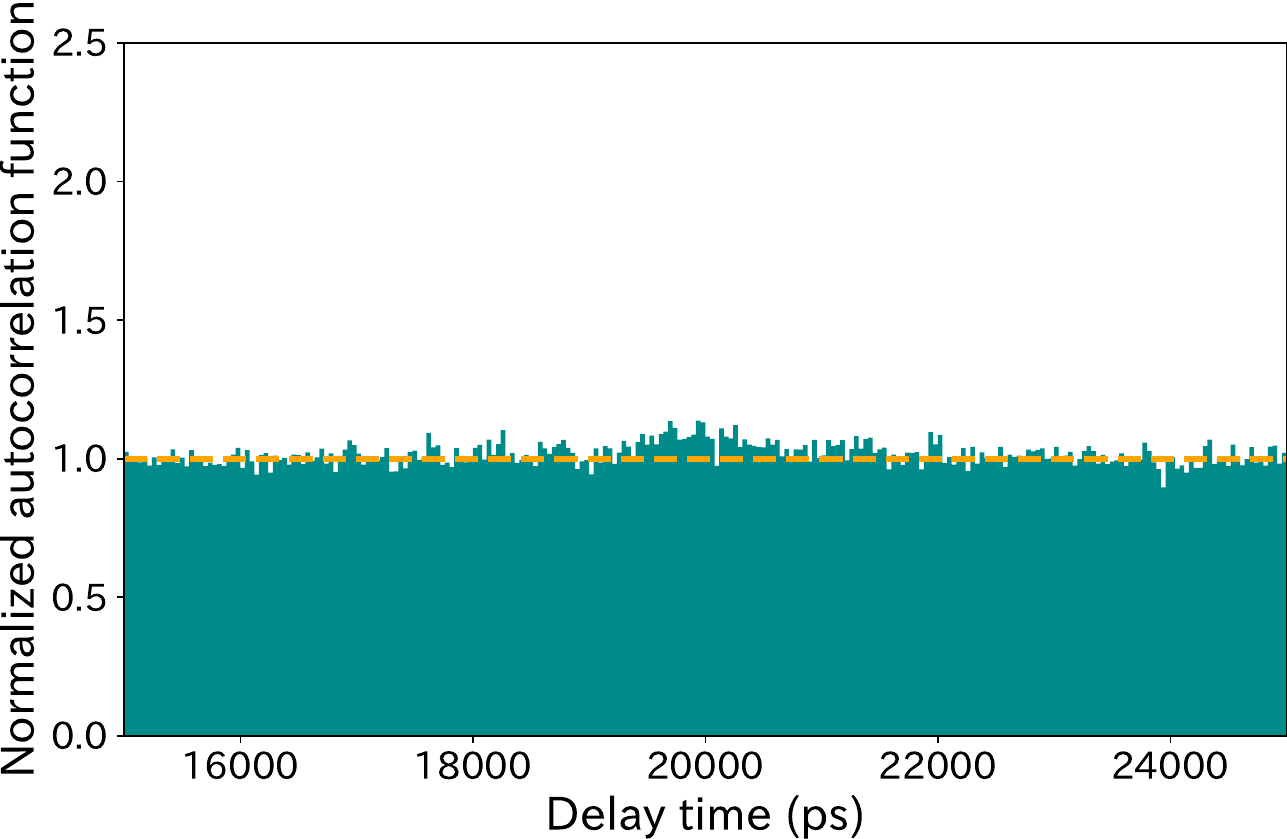}}
    \end{tabular}
    \caption{{\bf Temporal second-order autocorrelation.}
    The delayed coincidence counts between idler photons with 40-ps time resolution; this corresponds to the temporal second-order autocorrelation function.
    The polarizations are fixed in the horizontal direction.
    The central wavelength and bandwidth of the BPFs are respectively set to {\bf a} 1543.76~nm and 0.03 nm and {\bf b} 1543.89~nm and 0.30~nm. The data recording time is 1000~s. The orange lines represent the normalization factors estimated from the coincidence counts with a large delay. In {\bf a}, the bunching effect appears at the zero delay because unheralded photons behave like thermal light in the single-frequency mode. However, in {\bf b}, it is suppressed by the interference in multiple frequency modes. }
    \label{g2}
\end{figure*}

\begin{figure*}[tbp]
    \centering
    \begin{tabular}{cccc}
        \raisebox{-\height}{\bf a} & \raisebox{-1.1\height}{\includegraphics[scale=0.6, trim=0 0 0 0, clip]{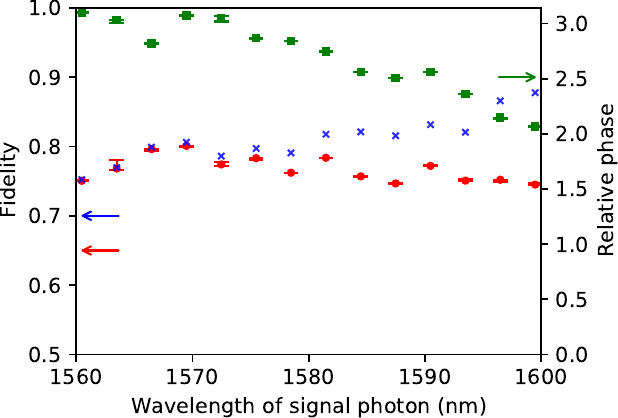}} &
        \raisebox{-\height}{\bf d} & \raisebox{-\height}{\includegraphics[scale=0.5, trim=0 0 0 0, clip]{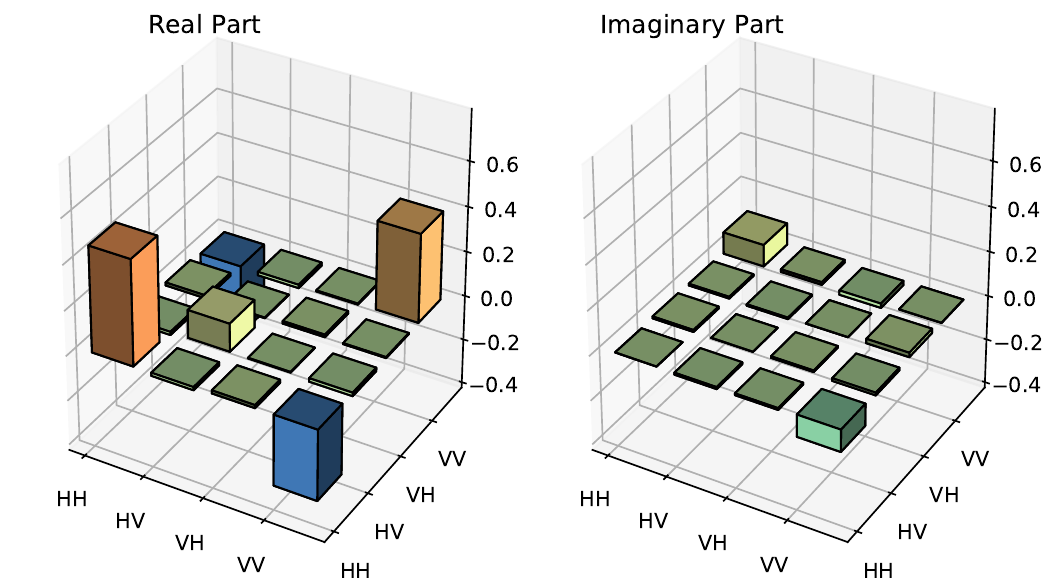}} \\
        \raisebox{-\height}{\bf b} & \raisebox{-1.1\height}{\includegraphics[scale=0.6, trim=0 0 0 0, clip]{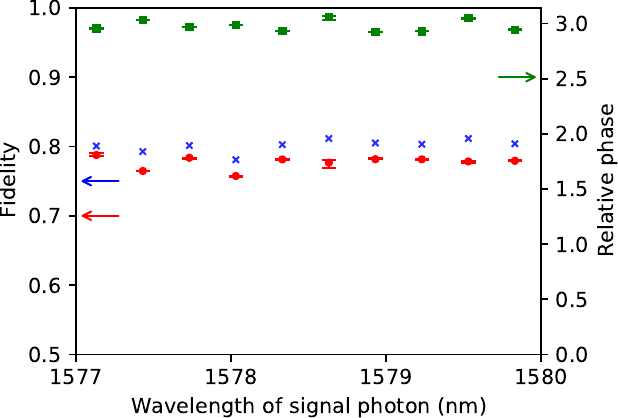}} &
        \raisebox{-\height}{\bf e} & \raisebox{-\height}{\includegraphics[scale=0.5, trim=0 0 0 0, clip]{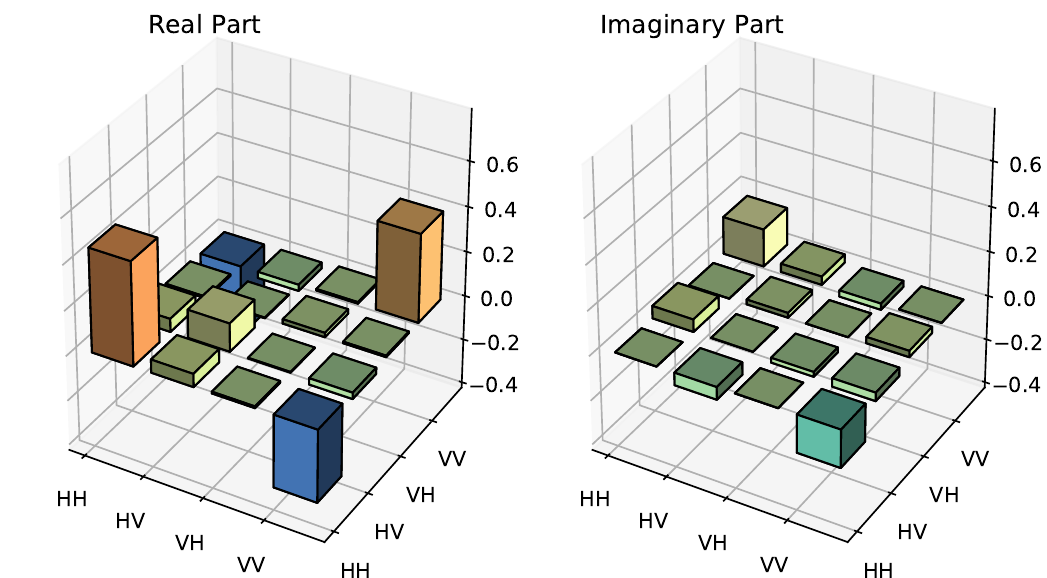}} \\
        \raisebox{-\height}{\bf c} & \raisebox{-1.1\height}{\includegraphics[scale=0.6, trim=0 0 0 0, clip]{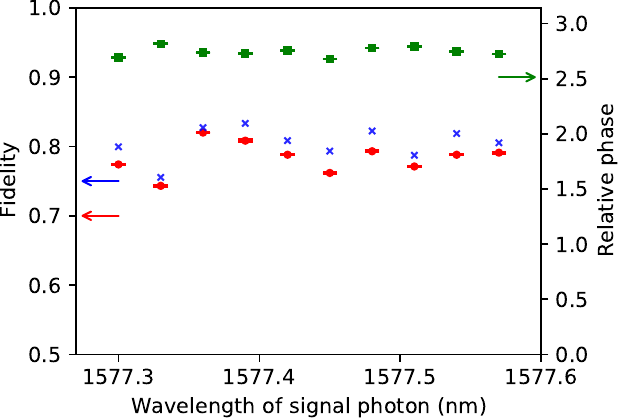}} &
        \raisebox{-\height}{\bf f} & \raisebox{-\height}{\includegraphics[scale=0.5, trim=0 0 0 0, clip]{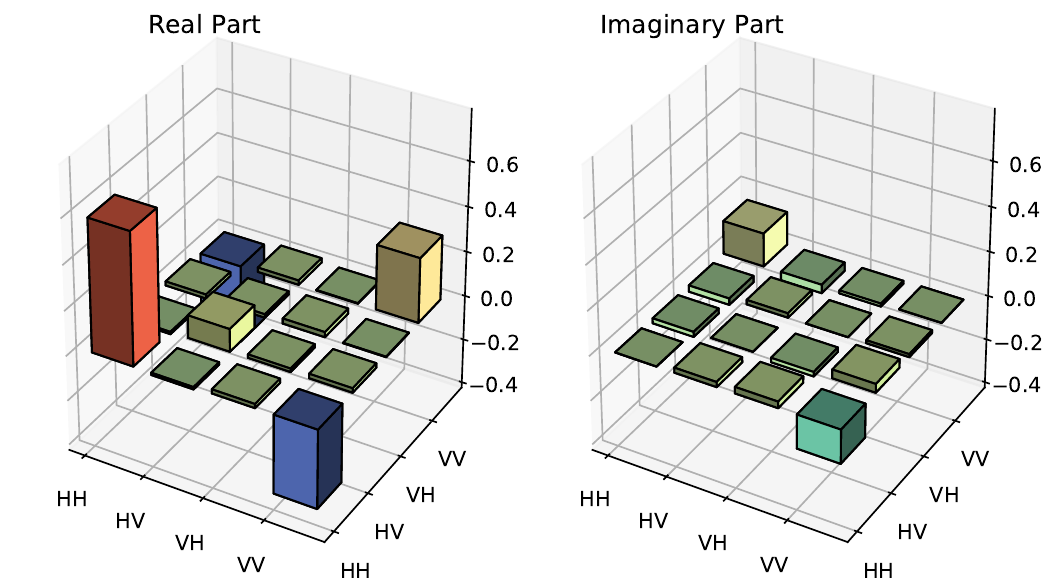}}
    \end{tabular}
    \caption{{\bf Fidelities and reconstructed density matrices of polarization entangled states.} We performed a set of polarization measurements with various central wavelengths and bandwidths of BPFs. The data recording time at each polarization setting is 200~s. {\bf a-c} Red, green, and blue markers represent the fidelity of the quantum state, relative phases between the polarizations of HH and VV of the reconstructed states, and expected fidelity after removing contamination with undesired components, respectively (see Supplementary Information). The error bars represent standard deviations under the assumption of Poisson statistics for the photon counts. The frequency ranges are {\bf a} [1560.48~nm, 1599.48~nm] and [1522.43~nm, 1560.48~nm], {\bf b} [1577.13~nm, 1579.83~nm] and [1541.60~nm, 1544.18~nm], and {\bf c} [1577.30~nm, 1577.57~nm] and [1543.76~nm, 1544.01~nm] for bandwidths of 3.00~nm, 0.30~nm, and 0.03~nm, which correspond to $\sim$ 100, 10, and 1 frequency modes, respectively. Examples of the reconstructed density matrix are shown in {\bf d}, {\bf e}, and {\bf f} with spectral resolutions of 3.00~nm, 0.30~nm, and 0.03~nm, respectively. The combinations of the frequency modes are {\bf d} 1578.48~nm and 1542.89~nm, {\bf e} 1577.43~nm and 1543.89~nm, and {\bf f} 1577.39~nm and 1543.93~nm.}
    \label{tomo}
\end{figure*}
The mode number contained in the dense and well-separated quantum frequency comb can be estimated using the temporal second-order autocorrelation function $g^{(2)}(\tau)$ of the signal (idler) photons. This method is well established in the pulsed pump SPDC, but not in the CW pump case. We derive the relation $\Delta g^{(2)} = (\pi\Delta f_{1/2}M)^{-1}$ for the CW-pumped SPDC with a singly resonant cavity by defining the time-integrated autocorrelation by $\Delta g^{(2)} = \int d\tau \qty(g^{(2)}(\tau)-1) $; it is similar to the relation between $g^{(2)}$ and the effective mode (Schmidt mode) for the pulsed-pump regime~\cite{Christ2012}. 
We obtain $\Delta g^{(2)} = 2.1$ and $0.17$ ns from the observed $g^{(2)}(\tau)$ with bandwidths of 0.03~nm and 0.30~nm, respectively, as shown in Figs.~4a and 4b.
We estimated $M$ to be 1.2 and 14.9 for bandwidths of 0.03 nm and 0.30 nm,
respectively, by using $\Delta f_{1/2}=126$ MHz. This is consistent with the mode numbers 1 and 10 expected from the beat frequency.

Finally, we perform the quantum state tomography of the polarization DOF~\cite{James2001} at various central frequencies. We adjust the pump power to 50~$\mu$W, 500~$\mu$W, and 5~mW for filter bandwidths of 3.00~nm, 0.30~nm, and 0.03~nm, respectively, where the excitation rates over a coincidence time window of 1100~ps are below 0.01. Examples of the matrix representations of the reconstructed density operators $\rho$ are shown in Fig.~5d-f. Fidelity is defined by $F(\rho) = \max_\theta\expval{\rho}{\Psi_\theta}$, where $\ket{\Psi_\theta} = \qty(\ket{H}\ket{H}+e^{i\theta} \ket{V}\ket{V})/\sqrt{2}$ is a maximally entangled state with a relative phase $\theta$. Figs.~5a-c show that the observed fidelities for all frequency regions and bandwidth settings are over 0.7, which clearly exceeds 0.5; this implies that quantum entanglement in the polarization subspace exists in all frequency ranges.
{See Fig. S5 in the Supplementary Information for other characterization of the generated states.}

\section{Discussion}
We now discuss how to improve the fidelity of the observed states. One of the reasons for the decrease in fidelity is contamination with undesired components $\ket{HV}$ and $\ket{VH}$ caused by the insufficient difference between the two path lengths on the Sagnac interferometer (Fig.~S3 in the Supplementary Information). The contamination can be prevented by placing the PPLN/WR in a more off-centered position of the Sagnac interferometer. Another solution is to place the PPLN/WR in the centered position and stabilizing the interferometer, which can not only improve the fidelity but also double the generation rate. The expected fidelities after removing these contaminants are shown in Figs.~5a-c by blue markers. However, there remains a non-negligible decrease in the fidelity. Although this can be attributed to experimental imperfections such as the insufficient temperature control of the PPLN/WR, further investigations are required in the future.

We developed polarization-entangled BFCs with over 1000 frequency modes using a singly resonant PPLN/WR in a Sagnac interferometer. We showed that the SPDC-based entangled BFC generator enables us to prepare both high-dimensional hyperentangled photon pairs and frequency-multiplexed entangled photon pairs by adjusting the pump power for the SPDC. The millimeter-long cavity structure realizes a comb spacing from gigahertz to tens of gigahertz, and therefore, commercially available filters and wavelength-division multiplexing devices can help realize a broadband quantum communication network.
In addition, the presented setup has the potential to integrate into one chip~\cite{Kotlyar2021,Elshaari2020}.
Thus, this versatile, stable, and highly efficient system for the generation of the massive-mode entangled BFCs can help provide a platform for high-capacity and highly efficient quantum information processing.

\section{Methods}
{\bf Singly Resonant PPLN/WR.} We used a 20~mm long-ridge waveguide that consists of a zinc-doped lithium niobate as the core and lithium tantalite as a clad (NTT Electronics), as a nonlinear optical waveguide. The periodic-poling period was designed to satisfy the type-0 quasi-phase-matching of the second harmonic generation of 1560.48 nm light at 35~$^\circ$C. Both end-faces of the waveguide were flat polished for the Fabry— P\'{e}rot cavity structure and coated by dielectric multilayers for a highly reflective coating around 1600~nm and anti-reflective coatings around 1520~nm and 780~nm~\cite{Ikuta2019d}. The waveguide resonator was stabilized by temperature control using a Peltier device.

{\bf Experimental Setup.} The pump light is generated by second-harmonic generation in PPLN/W pumped by an external cavity laser with a planar lightwave circuit (RIO Planex) at 1560.48~nm. The power and polarization of the pump light were adjusted using a pair of HWPs sandwiched between a PBS. We set the power of the pump light to 50~$\mu$W, 500~$\mu$W, and 5~mW to maintain the effective $M\zeta$ for different bandwidths of the BPFs, 3.00~nm, 0.30~nm, and 0.03~nm), respectively.

A BFC generated at the Sagnac interferometer was separated from the pump light by a dichroic mirror (DM), and it was divided into two paths by a BS with a probability of 1/2. Each photon was projected onto any polarization state using an HWP, a quarter-wave plate(QWP), and a PBS. The photon was coupled to a single-mode fiber connected to a tunable BPF (Alnair Labs), whose bandwidth could be adjusted between 0.03 and 3.00 nm. Finally, the photon was detected using a SNSPD~\cite{Miki2017}. The electrical signals from the two SNSPDs were used as the start and stop of a time-to-digital converter (PicoQuant) to record the coincidence counts with timestamps. For quantum state tomography, the coincidence counts were totaled in the time windows of 5800--6900~ps. 

{\bf Formula for Estimating the Number of Cavity Modes in the CW Pump Regime.} 
In the pulsed pump regime, the autocorrelation function of the signal or idler photons of photon pairs is related to the number of effective (Schmidt) modes by $g^{(2)} = 1+\frac{1}{N}$ under the assumption of a uniform spectral amplitude~\cite{Christ2012}. However, this formula cannot be applied directly to the CW-pumped regime. Autocorrelation is related to the cavity mode in the CW-pumped regime, which is the same as the effective mode only in the case of ideal frequency-bin entangled states. We obtained a formula 
\begin{align}
    \Delta g^{(2)} &\equiv \int d\tau \qty(g^{(2)} (\tau)-1) \notag \\
    &=\frac{1}{M} \frac{\gamma_i^2+3\gamma_i\gamma_s+\gamma_s^2}{\gamma_i \gamma_s(\gamma_i+\gamma_s)} \notag \\
    &\qty(=
        \begin{cases}
            \frac{5}{2\gamma_s M} & (\gamma_s=\gamma_i) \\
            \frac{1}{\gamma_s M} & (\gamma_s \ll \gamma_i)
        \end{cases}
    ),
    \label{auto_formula}
\end{align}

where $\gamma_{s(i)} = \pi \Delta f^{s(i)}_{1/2}$, and $\Delta f^{s(i)}_{1/2}$ represents the FWHM of the signal (idler) photons. The derivation is provided in the Supplementary Information.

\section{References}
\bibliography{sagnac.bib}
 
\begin{acknowledgments}
\section{Acknowledgements}
R.I., N.I., and T.Yamamoto acknowledge the members of the Quantum Internet Task Force for the comprehensive and interdisciplinary discussions on the Quantum Internet.
This work was supported by CREST, JST JPMJCR1671;
MEXT/JSPS KAKENHI Grant Numbers
JP20H01839, JP18K13483, and JP20J20261; Asahi Glass Foundation; and Program for Leading Graduate Schools: Interactive Materials Science Cadet Program.

\section{Author Contributions}
T. Yamazaki, R.I., and T.K. designed and performed the experiments and analyzed the data.
T. Yamazaki and R.I. led the theoretical analysis.
S.M., F.C., and H.T. developed the system of superconducting single-photon detectors.
T. Yamamoto and N.I. supervised and managed the project.
T. Yamazaki drafted the manuscript with inputs from all authors.

$\ $ 

\section{Data availability}
The experimental data and source code supporting the findings of this study are available from the corresponding author on reasonable request. 

\section{Competing interests}
The authors declare no competing interests. 

\end{acknowledgments}
\end{document}


\title{Supplementary Information for Massive-mode polarization entangled biphoton frequency comb} 

\author{Tomohiro Yamazaki}
\affiliation{Graduate School of Engineering Science, Osaka University, Toyonaka, Osaka 560-8531, Japan}

\author{Rikizo Ikuta}
\affiliation{Graduate School of Engineering Science, Osaka University, Toyonaka, Osaka 560-8531, Japan}
\affiliation{Center for Quantum Information and Quantum Biology, Osaka University, Osaka 560-8531, Japan}

\author{Toshiki Kobayashi}
\affiliation{Graduate School of Engineering Science, Osaka University, Toyonaka, Osaka 560-8531, Japan}
\affiliation{Center for Quantum Information and Quantum Biology, Osaka University, Osaka 560-8531, Japan}

\author{Shigehito Miki}
\affiliation{Advanced ICT Research Institute, National Institute of Information and Communications Technology (NICT), Kobe 651-2492, Japan}
\affiliation{Graduate School of Engineering, Kobe University, Kobe 657-0013, Japan}

\author{Fumihiro China}
\affiliation{Advanced ICT Research Institute, National Institute of Information and Communications Technology (NICT), Kobe 651-2492, Japan}

\author{Hirotaka Terai}
\affiliation{Advanced ICT Research Institute, National Institute of Information and Communications Technology (NICT), Kobe 651-2492, Japan}

\author{Nobuyuki Imoto}
\affiliation{Center for Quantum Information and Quantum Biology, Osaka University, Osaka 560-8531, Japan}

\author{Takashi Yamamoto}
\affiliation{Graduate School of Engineering Science, Osaka University, Toyonaka, Osaka 560-8531, Japan}
\affiliation{Center for Quantum Information and Quantum Biology, Osaka University, Osaka 560-8531, Japan} 

\date{\today}

\maketitle

\subsection{Analysis of state generated from the singly resonant PPLN/WR \\ and its second-order cross-correlation function}
We briefly summarize the theoretical analysis of photon pair generation by SPDC inside a cavity, called the cavity-enhanced SPDC~\cite{Ou1999,Lu2000,Herzog2008,Scholz2009a,Jeronimo-Moreno2010,Zielinska2014a,Luo2015,Couteau2018a}. Then, we explain the fitting function used for the analysis of the cross-correlation function of the photon pairs and its relationship with the spectral properties of the generated BFCs. A photon pair generated by a cavity-enhanced SPDC is represented by

\begin{align}
    \ket{\Psi} &= \int d\omega_i d\omega_s f(\omega_i, \omega_s) a^\dag(\omega_i) a^\dag(\omega_s) \ket{0} \notag \\
    f(\omega_i,\omega_s) &\propto \alpha_p(\omega_i+\omega_s) h(\Delta k) \mathcal{A} (\omega_i) \mathcal{A} (\omega_s).
    \label{state}
\end{align}

$f(\omega_i, \omega_s)$ represents the joint spectral amplitude (JSA) of the signal and idler photons, normalized by $\int d\omega_i d\omega_s \abs{f(\omega_i, \omega_s)}^2 =1$. $\alpha_p(\omega_i+\omega_s)$ denotes the spectral amplitude of the pump light. In the CW pump regime, the linewidth of the CW pump laser is sufficiently narrow, and we approximate it to $\alpha_p(\omega_i+\omega_s) \propto \delta(\omega_p-\omega_i-\omega_s)$. $h(\Delta k)=L \text{sinc}(L \Delta k /2)$ represents a function of the phase-matching condition, where $\Delta k(\omega_i,\omega_s) = n(\omega_i+\omega_s)(\omega_i+\omega_s)-n(\omega_i)\omega_i-n(\omega_s)\omega_s$, $L$ denotes the length of the nonlinear crystal, and $n$ denotes the refractive index\cite{ou2007multi}. $\mathcal{A} (\omega)$ is a function that represents the transmission spectrum of the cavity~\cite{Collett1985,walls2007quantum,garrison2008quantum}; it is represented as
\begin{equation}
    \mathcal{A} (\omega) \propto \sum_{m=0}^{\infty}\frac{\sqrt{\gamma(\omega)}}{\gamma(\omega)-i(\omega-m\Delta(\omega))}
\end{equation}
for a symmetric cavity without internal losses, where $\Delta(\omega)=\frac{\pi c }{n(\omega)L}$. $L$ denotes the length of the resonator; we assume that is equal to the length of the nonlinear crystal, and $c$ represents the speed of light. 
The resonant frequencies $\omega_m$ for integers $m$, which are the peaks of the observed spectrum, are determined by the solutions of the equation $\omega = m\Delta(\omega)$.
The FWHM is $\Delta f_{1/2} = \gamma/\pi$ and the FSR is $\frac{c}{2n_g L}$, where $n_g(\lambda_0)=n(\lambda_0)-\lambda_0 \frac{dn}{d\lambda_0}$ and $\lambda_0$ represent the group index and the wavelength in vacuum, respectively.

\nolinenumbers
\begin{figure}[tbp]
    \centering
    \includegraphics[scale=0.7, trim=0 0 0 0, clip]{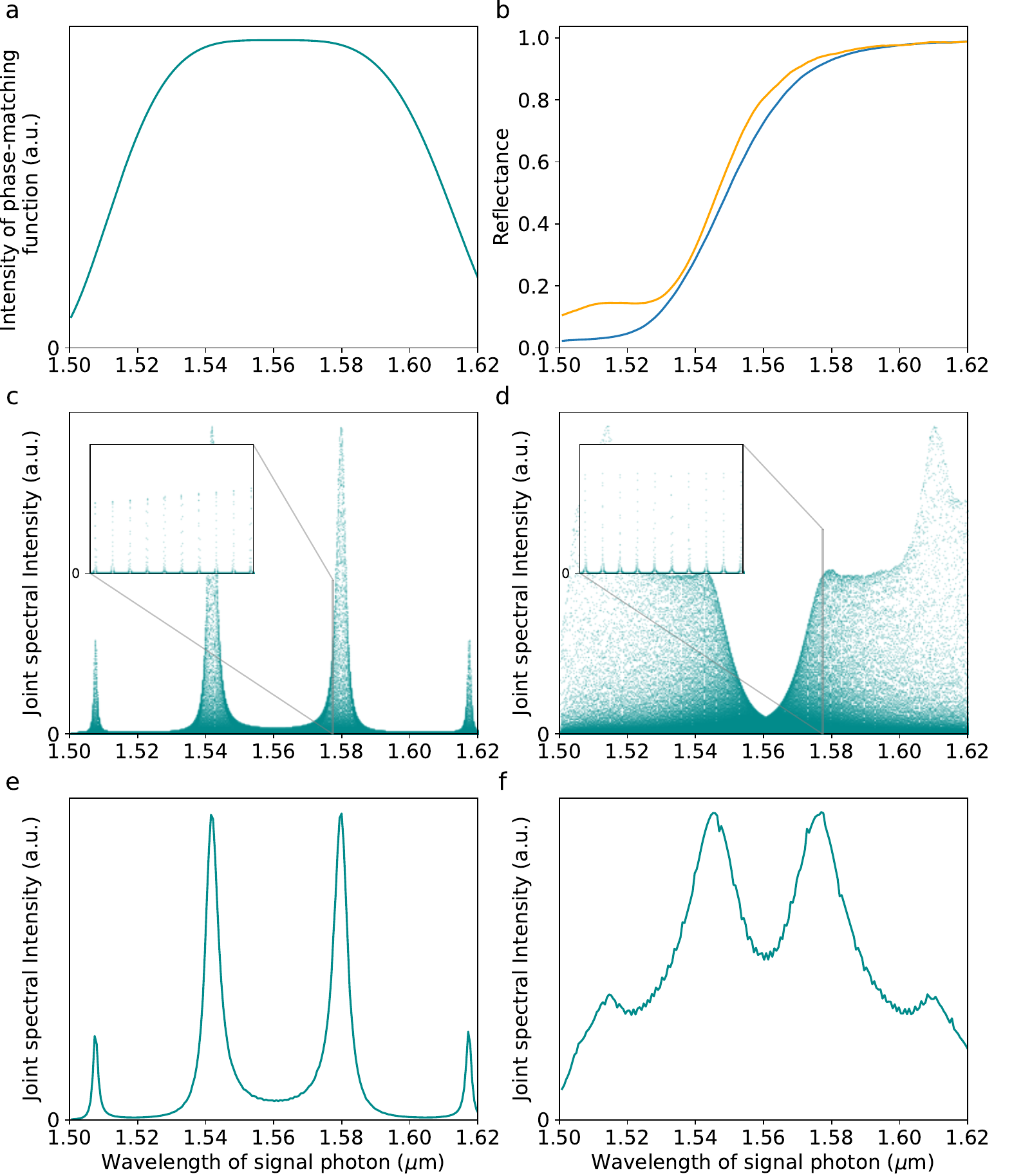}
    \caption{{\bf Simulation of the frequency spectra of biphoton frequency combs.} {\bf a} Phase-matching function $|h(\Delta k)|^2$. The refractive indices are calculated from Ref.~\cite{Jundt1997}, where we conveniently set the periodic polling period to satisfy $\Delta k =0$ at 1560.48~nm and the temperature to $T=69.95 ^\circ \mathrm{C}$. {\bf b} Frequency dependence of the reflectance of our waveguide resonator. The two curves represent the reflectances of the front and back-end faces. {\bf c} Frequency spectrum of biphoton frequency comb in doubly resonant configuration. We set the reflectance $R_{r(l)}=\abs{r_{r(l)}}^2=0.9$ and $\gamma = -c\log{(r_r r_l)}/2n(\omega)L$ and neglected the internal loss. {\bf d} Frequency spectrum of the biphoton frequency comb in the singly resonant configuration. The insets in {\bf b} and {\bf c}, whose frequency ranges are [1577.30~nm, 1577.57~nm], show the periodic peaks at the resonant frequencies.   {\bf e} ({\bf f}) Results after integrating the JSI of {{\bf c} ({\bf d})} over a frequency range of 0.3~nm .
    }
    \label{simulation}
\end{figure}
The temperature-dependent Sellmeier equation described in Ref.~\cite{Jundt1997} for the rough estimation of the refractive index $n(\lambda_0)$ is used to calculate the phase-matching function $|h(\Delta k)|^2$ shown in Fig.~S1a.
For the doubly resonant configuration, where both signal and idler photons are confined by the cavity, photon pair generation in most frequency ranges is suppressed because of the frequency dependency of the FSR, which is known as the ``cluster effect"~\cite{Pomarico2012}.
Figure~S1c shows the simulation of the spectrum of the biphoton frequency comb in the doubly resonant configuration from equation \eqref{state}, where the cluster effect appears clearly.

We realized a singly resonant configuration using mirrors whose reflectance is frequency-dependent, as shown in Fig.~S1b.
Using these values of the reflectance, we simulate the spectrum of the biphoton frequency comb in our experiment, as shown in Fig.~S1d.
The biphoton frequency comb is generated with a certain intensity in all frequency ranges.
The cluster effect does not appear if the singly resonant configuration is completely realized.
However, this is not realized around the degenerate point of 1560.48~nm in our experiments.
The cluster effect around the degenerate point causes a nonuniform shape in the spectrum of the biphoton frequency comb.
{The raw joint spectral intensities in Fig.~S1c and S1d are very sensitive to the frequency difference because of their fine comb shape.
Thus, we need to integrate these in the bandwidth for the measurement of 0.3~nm for comparison with the measurement results of the JSI. 
Figs.~S1e and S1f show the results after integrating Fig.~S1c and S1d, in this range, respectively.}

We further simplified JSA $f(\omega_i, \omega_s)$ 
to analyze the cross-correlation measurements.
The frequency windows of these measurements in our experiment are up to 3~nm, which is sufficiently small to neglect the frequency dependency of $n(\omega)$ and $\gamma(\omega)$ and the cluster effect within its frequency windows.
Therefore, we replace $m\Delta$ by a resonant frequency $\omega_m$.
When we set the pump frequency to satisfy $\omega_{m}+\omega_{m'} \simeq \omega_p$ and represent $\gamma(\omega_{s(i)})$ by $\gamma_{s(i)}$, we have

\begin{align}
    f(\omega_i, \omega_s) &\propto \delta(\omega_p-\omega_i-\omega_s) \sum_{m,m'} \frac{1}{\gamma_i-i\qty(\omega_i-\omega_m)} \frac{1}{\gamma_s-i\qty(\omega_s-\omega_{m'})} \notag \\
    &=\delta(\omega_p-\omega_i-\omega_s) \sum_{m} \frac{1}{\gamma_i-i\qty(\omega_i-\omega_m)} \frac{1}{\gamma_s+i\qty(\omega_i-\omega_m)}.
    \label{jsa_ap}
\end{align}

In the last equality, only the term about $m'$ that satisfies $\omega_{m'}=\omega_p-\omega_m$ is left. The case $\gamma_i=\gamma_s$ corresponds to the (equally) doubly resonant configuration, and the case $\gamma_i \rightarrow \infty $ corresponds to the singly resonant configuration, where only signal photons are confined.

In the measurement of the coincidence counts between the signal and idler photons with a finite temporal resolution, the observed cross-correlation $C_{s, i}$ is described by 

\begin{equation}
    C^{(1,1)}_{s,i} (\tau) \propto \int d\tau' g_{\sqrt{2}\sigma}(\tau-\tau') G^{(1,1)}_{s,i}(\tau'),
    \label{C}
\end{equation}

where

\begin{equation}
    G^{(1,1)}_{s,i}(\tau) \equiv \expval{a_s(t)^\dag a_i(t+\tau)^\dag a_i(t+\tau) a_s(t)}{\Psi}
\end{equation}

is a temporal second-order cross-correlation function, and
\begin{equation}
    g_\sigma(t) = \frac{1}{\sqrt{2\pi}\sigma} \exp{-\frac{t^2}{2\sigma^2}}
\end{equation}
is a Gaussian function that approximates the finite temporal resolution of the two photon detectors used for coincidence measurement.~\cite{Herzog2008}. 

From equations~\eqref{state},~\eqref{jsa_ap}, and~\eqref{C},

\begin{align}
    C^{(1,1)}_{s,i} (\tau) &\propto \int d\tau' g_{\sqrt{2}\sigma}(\tau-\tau') \abs{\sum_m \int d\omega_s \frac{e^{-i\qty(\omega_p-\omega_s)\tau'}}{\qty{\gamma_i+i\qty(\omega_s-\omega_m)}\qty{\gamma_s-i\qty(\omega_s-\omega_m)}}}^2 \\
    &\propto \int d\tau' g_{\sqrt{2}\sigma}(\tau-\tau') \qty[e^{-2\gamma_i\tau'} \abs{\sum_m{e^{i\omega_m\tau'}}}^2 h(\tau')+
    e^{2\gamma_s\tau'} \abs{\sum_m{e^{i\omega_m\tau'}}}^2 h(-\tau')].
    \label{C_cal}
\end{align}

$h(\tau)$ represents the step function~\cite{Luo2015}. Equation~\eqref{C_cal} becomes when we extract a single-frequency mode using BPFs:

\begin{align}
    C^{(1,1)}_{s,i,\text{single}} (\tau) &\propto \int d\tau' g_{\sqrt{2}\sigma}(\tau-\tau') \qty(e^{-2\gamma_i\tau'} h(\tau')+e^{2\gamma_s\tau'} h(-\tau')) \notag \\
    &\propto \frac{1}{2} e^{\qty(2\gamma_i \sigma)^2 - 2\gamma_i \tau} \qty{1 - \erf\qty(-\frac{\tau}{2\sigma} + 2\gamma_i\sigma)} + \frac{1}{2} e^{\qty(2\gamma_s \sigma)^2 + 2\gamma_s \tau} \qty{1 + \erf\qty(-\frac{\tau}{2\sigma} - 2\gamma_s\sigma)},
    \label{single}
\end{align}

where $\erf(x)$ denotes an error function. When we extract many frequency modes enough to the approximation, 

\begin{equation}
    \abs{\sum_m{e^{i\omega_m\tau'}}} = \abs{\sum_m{e^{2\pi i m \frac{\tau'}{T_0}}}} \propto \sum_n \delta(\tau'-nT_0),
\end{equation}

where $T_0 = 2\pi/\text{FSR}$ represents the round-trip time of the cavity.
Equation~\eqref{C_cal} then becomes

\begin{equation}
    C^{(1,1)}_{s,i,\text{multi}} (\tau) \propto \frac{1}{2\sqrt{\pi}\sigma} \qty[ \exp(\frac{-\tau^2}{4\sigma^2}) +\sum_{j=1}^{\infty} \qty[\exp(-2\gamma_i j T_0 -\frac{(jT_0-\tau)^2}{4\sigma^2}) +\exp(-2\gamma_s j T -\frac{(jT_0+\tau)^2}{4\sigma^2})] ].
     \label{multi}
\end{equation}

Coincidence counts are represented by equation~\eqref{single} if the state is a statistical mixture of the photon pairs of the different frequency modes. The cross-correlation can be described by considering a mixture of equations~\eqref{single} and \eqref{multi} with a normalization factor with relative weight $p$ as

\begin{equation}
    C^{(1,1)}_{s,i,\text{sum}} (\tau) \propto p \frac{(1-e^{-2\gamma_i T})(1-e^{-2\gamma_s T})}{1-e^{-2\qty(\gamma_i+\gamma_s) T}} C^{(1,1)}_{s,i,\text{multi}} (\tau) + \qty(1-p) \frac{2\gamma_i \gamma_s}{\gamma_i+\gamma_s} C^{(1,1)}_{s,i,\text{single}} (\tau).
    \label{sum}
\end{equation}

In our experiment, the best fit of this function in equation~\eqref{sum} to the observed temporal second-order cross-correlation in Fig.~S4 leads to $p \simeq 1$.

\subsection{Analysis of the state generated from the Sagnac interferometer and its fidelity}
\nolinenumbers
\begin{figure}[tbp]
    \centering
    \begin{tabular}{cccc}
        \raisebox{-\height}{\bf a} & \raisebox{-\height}{\includegraphics[scale=1, trim=0 0 0 0, clip]{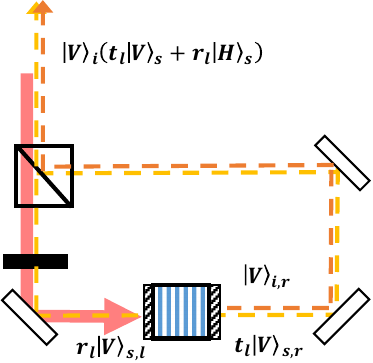}} &
        \raisebox{-\height}{\bf b} & \raisebox{-\height}{\includegraphics[scale=1, trim=0 0 0 0, clip]{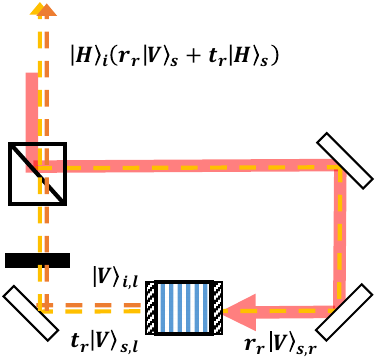}}
    \end{tabular}
    \caption{{\bf Sagnac interferometer with cavity-enhanced spontaneous parametric down-conversion.} The thick red arrow represents the propagation of the pump laser and the thin yellow (orange) arrow represents the propagation of the signal (idler) photons. The subscript r (l) indicates the photon exiting the right (left) side of PPLN/WR.}
    \label{sagnac}
\end{figure}
We considered only one photon-pair state generated by the SPDC. As shown in Fig.~S2, owing to the cavity structure for the signal photons, H-polarized and V-polarized pump lights respectively generate photon pairs described by

\begin{equation}
    g_H e^{i\qty(\omega_p\tau_l+\theta(\omega_p))}\qty({t_l\ket{V}_{i,r}\ket{V}_{s,r}+r_l\ket{V}_{i,r} \ket{V}_{s,l}}),
\end{equation}

\begin{equation}
    g_V e^{i\omega_p\tau_r}\qty({t_r\ket{V}_{i,l} \ket{V}_{s,l}+r_r\ket{V}_{i,l}\ket{V}_{s,r}}),
\end{equation}

where $g_{H(V)}$, $\theta(\omega)$, and $\tau_{l(r)}$ represent the coupling constant proportional to the complex amplitude of the H (V)-polarized pump light, frequency-dependent phase from a multiorder HWP, and propagation time of the left (right) arm of the Sagnac interferometer in Fig.~S2, respectively. $t_{l(r)}$ and $r_{l(r)}$ represent the transmittance and reflectance of the PPLN/WR when the light enters from the left (right) side, respectively, where $\abs{t_{l(r)}}^2+\abs{r_{l(r)}}^2=1$ holds. The output state just before photon detection is represented by

\begin{widetext}

    \begin{align}
        \ket{\Psi} &\propto g_{H}\eta_{ir}e^{i\qty(\omega_p\tau_l+\theta(\omega_p)+\omega_i\tau_r)}\qty(t_l\eta_{sr}e^{i\qty(\omega_s\tau_r)}\ket{V}_{i}\ket{V}_{s}+r_l\eta_{sl}e^{i\qty(\omega_s\tau_l+\theta(\omega_s))}\ket{V}_{i}\ket{H}_{s}) \notag \\
        &+g_{V}\eta_{il}e^{i\qty(\omega_p\tau_r+\omega_i\tau_l+\theta(\omega_i))}\qty(t_r\eta_{sl}e^{i(\omega_s\tau_l+\theta(\omega_s))}\ket{H}_{i}\ket{H}_{s}+r_r\eta_{sr}e^{i\omega_s\tau_r}\ket{H}_{i}\ket{V}_{s}) \notag \\
        &\propto g_{H}\eta_{ir}\ket{V}_{i}\qty(t_l\eta_{sr}\ket{V}_{s}+r_l\eta_{sl}e^{-i\Delta \theta}\ket{H}_{s}) +g_{V}\eta_{il}e^{i\phi}\ket{H}_{i}\qty(t_r\eta_{sl}\ket{H}_{s}+r_r\eta_{sr}e^{i\Delta \theta}\ket{V}_{s}), \\
        &\phi = \theta(\omega_s) +\theta(\omega_i) -\theta(\omega_p), \quad \Delta \tau = \tau_r-\tau_l, \quad \Delta \theta = \omega_s\Delta\tau-\theta(\omega_s),
    \end{align}

\end{widetext}
where $\eta_{i(j)l(r)}$ represents the total loss in the optical circuit for the signal (idler) photons coming from the left (right) side of the PPLN/WR. By setting the polarization of the pump light to satisfy $g_{H}\eta_{ir}t_l\eta_{sr}=g_{V}\eta_{il}t_r\eta_{sl}$, we obtain

\begin{equation}
    \ket{\Psi} \propto \ket{V}_{i}\qty(\ket{V}_{s}+\beta_H e^{-i\Delta \theta}\ket{H}_{s}) 
    +e^{i\phi}\ket{H}_{i}\qty(\ket{H}_{s}+\beta_V e^{i\Delta \theta}\ket{V}_{s}),
    \label{4output}
\end{equation}

where

\begin{equation}
    \beta_H \equiv \frac{r_l\eta_{sl}}{t_l\eta_{sr}}, \qquad \beta_V \equiv \frac{r_r\eta_{sr}}{t_r\eta_{sl}}.
\end{equation}

\nolinenumbers
\begin{figure}[tbp]
    \centering
    \begin{tabular}{cc}
        \raisebox{-\height}{\bf a} & \raisebox{-\height}{\includegraphics[scale=1, trim=0 0 0 0, clip]{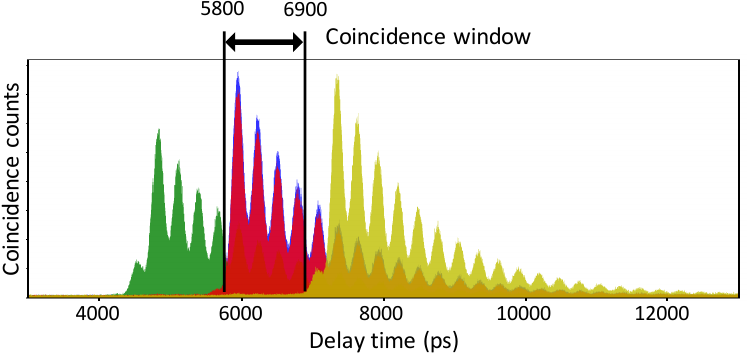}} \\
        \raisebox{-\height}{\bf b} & \raisebox{-\height}{\includegraphics[scale=0.8, trim=0 0 0 0, clip]{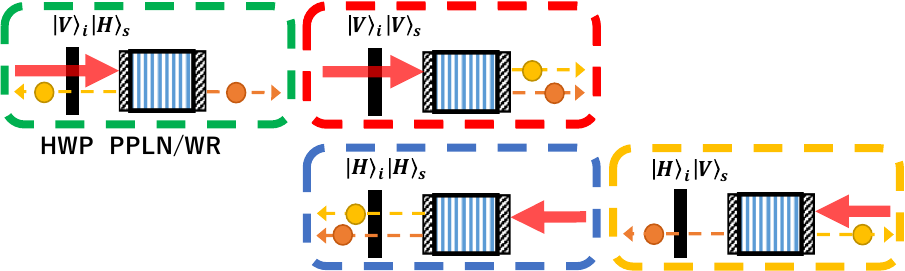}}
    \end{tabular}
    \caption{{\bf Correspondence of the arrival time, path, and polarization of the four output states.} {\bf a} Experimental data of coincidence counts in four different polarization settings, HH (blue), VV (red), VH (green), and HV (yellow), at wavelengths of 1580.48~nm and 1540.98~nm. VH (HV) photon pairs are detected earlier (later) than the VV and HH photon pairs. {\bf b} The paths of the polarizing photons in the Sagnac interferometer (see Fig.~S2). The difference in the arrival times is attributed to the difference in propagation paths.}
    \label{corresp}
\end{figure}

For $\Delta \tau \gg 0$, we obtain the maximally polarization-entangled state $\ket{\Phi} \equiv 1/\sqrt{2} \qty(\ket{H}_i\ket{H}_s+e^{i\phi}\ket{V}_i\ket{V}_s)$. However, in our experiment, there is an overlap between $\ket{V}_i\ket{H}_s$ and $\ket{\Phi}$, as indicated in Fig.~S3. To estimate these effects on the fidelity, we consider the delay time dependency of the generated state as 

\begin{align}
    \ket{\Psi (\tau)} &\propto 
    \ket{V,t}_{i}\qty(e^{-\gamma \tau}h(\tau)\ket{V,t+\tau}_{s}+\beta_H e^{-i\Delta \theta}e^{-\gamma(\tau+\Delta\tau)} h(\tau+\Delta\tau)\ket{H,t+\tau}_{s}) \notag \\
    &+e^{i\phi}\ket{H,t}_{i}\qty(e^{-\gamma \tau}h(\tau)\ket{H,t+\tau}_{s}+\beta_V e^{i\Delta \theta}e^{-\gamma(\tau-\Delta\tau)} h(\tau-\Delta\tau)\ket{V,t+\tau}_{s}).
\end{align}

Here, we approximated the temporal shaping of the state by $e^{-\gamma \tau}$.
In our experiment, we selected the time windows from 0 to $\Delta \tau $, i.e., 

\begin{align}
    \ket{\overline{\Psi}} \equiv \int_{0}^{\Delta\tau} d\tau \ket{\Psi (\tau)} \propto \ket{\Phi}+\frac{\beta_H}{\sqrt{2}} e^{-\gamma\Delta\tau-i\Delta\theta}\ket{V}_i\ket{H}_s.
\end{align}

We redefined this as

\begin{equation}
    \ket{H(V)}_i \ket{H(V)}_s \equiv \int_{0}^{\Delta\tau} e^{-\gamma \tau} \ket{H(V),t}_i \ket{H(V),t+\tau}_{s}.
\end{equation}

Further, the randomization of the phase of the component $\ket{V}_i\ket{H}_s$ leads to

\begin{equation}
    \rho_{\overline{\Psi}} \propto \qty(\ket{\Phi}\bra{\Phi} + \frac{\abs{\beta_H}^2}{2} e^{-2\gamma\Delta\tau} \ket{VH}\bra{VH}_{is}).
\end{equation}

From the definition of fidelity, $F(\rho) \equiv \max_\theta \expval{\rho}{\Psi_\theta}$, where $\ket{\Psi_\theta} \equiv 1/\sqrt{2} \qty(\ket{HH}+e^{i\theta}\ket{VV})$, 
\begin{equation}
    F(\rho_{\overline{\Psi}}) = \qty(1+ \frac{\abs{\beta_H}^2}{2}e^{-2\gamma\Delta\tau})^{-1}
\end{equation}
holds. From these results, we estimated the fidelity expected in the case of $\Delta \tau \gg 0$ by multiplying $F(\rho_{\overline{\Psi}})^{-1}$ with the experimentally measured fidelity (Fig.~5 in the manuscript). From the best fit to the data of the temporal cross-correlation in every polarization and frequency range, we extracted $\gamma$ and $\beta_H$, which are estimated as the ratio of the amplitude between the VH and HH components.
 
\subsection{Analysis of the generated polarization entangled BFC}
In the manuscript, we discuss a simplified model for generating polarization-entangled BFCs. Here, we analyze the polarization-entangled BFCs generated in the experiment using a more realistic model. As well as equation~\eqref{state}, the time evolution of the cavity-enhanced SPDC for arbitrary pump power is represented by

\begin{equation}
    \ket{\Psi} \propto \exp[g_{H(V)} \int d\omega_i d\omega_s f_{H(V)}(\omega_i, \omega_s) a_{H(V)}^\dag(\omega_i) a_{H(V)}^\dag(\omega_s) - h.c.] \ket{0} \equiv U_{H(V)} \ket{0}.
    \label{spdc}
\end{equation}

Equation~\eqref{spdc} is reduced to approximately equation~\eqref{state} when $\abs{g}^2 \ll 1$ is satisfied. In the CW pump regime, the JSA $f_{H(V)}(\omega_i, \omega_s)$ cannot be decomposed like $\sum_m c_{H(V),m} f_{H(V),m} (\omega_i) f'_{H(V),m} (\omega_s)$. Instead, we decompose JSA as 

\begin{equation}
    f_{H(V)}(\omega_i, \omega_s) \propto \delta(\omega_p-\omega_i-\omega_s) \sum_m c_{H(V),m} f_{H(V),m} (\omega_i).
\end{equation}

$c_{H(V),m}$ represents the coefficient of each cavity mode and $f_{H(V),m} (\omega_i)$ is normalized by $\int d\omega_i \abs{f_{H(V),m}(\omega_i)}^2 = 1$. We assume that each frequency mode is well separated from the others, that is, $\int d\omega_i f_{H(V),m}^*(\omega_i) f_{H(V),m'}(\omega_i) \simeq \delta_{m,m'}$. The generation processes of each cavity mode are independent.

\begin{equation}
    U_{H(V)} \ket{0} \simeq \prod_m \exp[g_{H(V)} c_{H(V),m} \int d\omega_i f_{H(V),m} (\omega_i) a_{H(V)}^\dag(\omega_i) a_{H(V)}^\dag(\omega_p-\omega_i) - h.c.]\ket{0} \equiv \prod_m U_{H(V),m} \ket{0}.
\end{equation}

Modifying equation~\eqref{4output} for an arbitrary pump power using the above results, the state generated by our experiments can be represented as
\begin{equation}
    \ket{\Psi} = \prod_m U_{H,m} U_{V,m} \ket{0},
    \label{output}
\end{equation}
where we ignored the component of the reflected signal photons. 
In the case of $\abs{g_H c_{H,m}}^2 + \abs{g_V c_{V,m}}^2 \ll 1$, equation~\eqref{output} is approximately

\begin{align}
    \ket{\Psi} &\simeq \bigotimes_m \frac{1}{\sqrt{1+\abs{g_H c_{H,m}}^2 + \abs{g_V c_{V,m}}^2}} \notag \\
    &\qty(1 + g_H c_{H,m} \int d\omega_i f_{H,m}(\omega_i) a_H^\dag(\omega_i) a_H^\dag (\omega_p-\omega_i).
    + g_V c_{V,m} \int d\omega_i f_{V,m}(\omega_i) a_V^\dag(\omega_i) a_V^\dag (\omega_p-\omega_i))
    \ket{0}_m.
\end{align}

This state can be considered a frequency-multiplexed entangled photon pair, which is similar to the state in equation (3) in the manuscript.
When $\sum_m (\abs{g_H c_{H,m}}^2 + \abs{g_V c_{V,m}}^2) \ll 1$, the state is approximately

\begin{align}
    \ket{\Psi} &\simeq \frac{1}{\sqrt{1+\sum_m(\abs{g_H c_{H,m}}^2 + \abs{g_V c_{V,m}}^2)}} \notag \\
    &\qty(1 + \sum_m \qty(g_H c_{H,m} \int d\omega_i f_{H,m}(\omega_i) a_H^\dag(\omega_i) a_H^\dag (\omega_p-\omega_i),
    + g_V c_{V,m} \int d\omega_i f_{V,m}(\omega_i) a_V^\dag(\omega_i) a_V^\dag (\omega_p-\omega_i)))\ket{0}.
\end{align}

After removing the vacuum state, this state may correspond to that in equation~(2) in the manuscript.
The difference between these states discussed here and the simplified model in the manuscript is that the spectral correlation between the signal and idler photons remains in each mode in these states; this often has unfavorable effects on the interference between independent photon pairs.

We have to use a pulsed-pump regime to generate the exact frequency-bin entangled state, which is expressed as $M^{-1/2} \sum_{m=1}^M \ket{\omega_{i,m}}\ket{\omega_{s,m}}$. This is approximately attained in the doubly resonant configuration by making the FWHMs of the cavities for the signal and idler photons considerably smaller than the FWHM of the pump laser
 because the frequencies of the signal and idler photons are determined independently and almost exclusively by the FWHMs of the cavities, respectively. 
In the singly resonant configuration, the exact frequency-bin entangled state itis also approximately attained by making the FWHM of the cavity for the signal photon considerably smaller than the FWHM of the pump laser.
Then, the frequencies of the signal and idler photons are determined almost exclusively by the FWHMs of the cavity and pump laser, respectively.
In this case, the FWHMs of both photons would differ significantly from each other.

\subsection{Theory of the autocorrelation function and the cavity mode in CW pump regime}
Ref.\cite{Christ2012} reported that a value $g^{(2)}$ relevant to $g^{(2)}(\tau)$ of the signal or idler photon of a photon pair in the pulse pump regime is related to the number of effective modes of a photon pair. The effective mode indicates the Schmidt mode of continuous variable Schmidt decomposition~\cite{Parker2000,Law2000a}. Under the assumption of a uniform spectral amplitude, the relationship between $g^{(2)}$ and the number of effective modes $N$ is expressed as $g^{(2)}=1 + \frac{1}{N}$. A similar discussion is held in the CW pump regime~\cite{Luo2015}, wherein $g^{(2)}(0) \simeq 1 + \frac{1}{M}$ is derived with the assumption that detectors have a finite temporal resolution, where $M$ represents the number of cavity modes. These results yield the same formula and have been applied in many experiments using pulse lasers~\cite{Harder2013,Kues2017} and CW lasers laser~\cite{Fortsch2013c,Shi2019,Seri2019}. Here, we present another formula for the CW pump regime, which is independent of the temporal resolution of the detectors.

From equation~\eqref{jsa_ap}, we define $f(\omega_i)$ and $f_m(\omega_i)$ as

\begin{equation}
    f(\omega_i, \omega_s) \equiv \delta(\omega_p-\omega_i-\omega_s) f(\omega_i)
    \equiv \delta(\omega_p-\omega_i-\omega_s) \frac{1}{\sqrt{M}} \sum_{m=1}^{M}  f_m(\omega_i)
    \label{multi_jsa},
\end{equation}

$f(\omega_i)$ and $f_m(\omega_i)$ are normalized. $f_m(\omega_i)$ represents the normalized spectral amplitude of each cavity mode, and we assume that each cavity mode is orthogonal.
We calculate the second-order autocorrelation function of the idler photons of the photon pairs as

\begin{equation}
    G^{(2)}_{i,i}(\tau) = \expval{a_i(t)^\dag a_i(t+\tau)^\dag a_i(t+\tau) a_i(t)}{\Psi}.
\end{equation}

Using the second-order term in equation~\eqref{spdc} without polarization dependency as the lowest-order approximation for $\ket{\Psi}$, we obtain

\begin{align}
    &G^{(2)}_{i,i} (\tau) \notag \\
    &\propto \norm{a_i(t+\tau) a_i(t) \int d\omega_{i1} \int d\omega_{i2} f(\omega_{i1}) f(\omega_{i2}) a_i^\dag(\omega_{i1}) a_s^\dag(\omega_p-\omega_{i1}) a_i^\dag(\omega_{i2}) a_s^\dag(\omega_p-\omega_{i2}) \ket{0}}^2 \notag \\
    &= \norm{\frac{1}{2\pi} \int d\omega_{i1} \int d\omega_{i2} f(\omega_{i1}) f(\omega_{i2}) \qty(e^{i\omega_{i1}\tau}+e^{i\omega_{i2}\tau}) a_s^\dag(\omega_p-\omega_{i1}) a_s^\dag(\omega_p-\omega_{i2}) \ket{0}}^2 \notag \\
    &=\frac{2}{\qty(2\pi)^2} \int d\omega_{i1} \int d\omega_{i2} \abs{f(\omega_{i1})}^2 \abs{f(\omega_{i2})}^2 \qty(2+e^{i(\omega_{i1}-\omega_{i2})\tau}+e^{-i(\omega_{i1}-\omega_{i2})\tau}) \notag \\
    &=\frac{1}{\pi^2} \qty[\abs{\int d\omega_i \abs{f(\omega_i)}^2}^2 + \abs{\int d\omega_i \abs{f(\omega_i)}^2 e^{-i\omega_i \tau}}^2] \notag \\
    &=\frac{1}{\pi^2} \qty[1 + 2\pi \abs{\mathcal{F}\qty[\abs{f(\omega_i)}^2] (\tau)}^2].
    \label{auto_f}
\end{align}

where $\mathcal{F}\qty[\abs{f(\omega_i)}^2]$ represents the Fourier transform of the spectral intensity of the idler photon.
When we define the normalized autocorrelation function as $g^{(2)}_{i,i}(\tau) \equiv G^{(2)}_{i,i}(\tau)/G^{(2)}_{i,i}(\tau \rightarrow \infty)$, we obtain

\begin{equation}
    g^{(2)}_{i,i} (\tau) = 1 + 2\pi \abs{\mathcal{F}\qty[\abs{f(\omega_i)}^2] (\tau)}^2.
    \label{n_auto}
\end{equation}

This result shows that $g^{(2)}_{i,i} (0) = 2$, which is independent of the spectral characteristics of photon pairs.
In the single mode case~\cite{Luo2015}, equation~\eqref{n_auto} is calculated as

\begin{equation}
    g^{(2)}_{i,i}(\tau) = 1 + \abs{\frac{1}{\gamma_i-\gamma_s}\qty(\gamma_i e^{-\gamma_s\abs{\tau}}-\gamma_s e^{-\gamma_i\abs{\tau}})}^2.
    \label{single_auto}
\end{equation}

In the multimode case, $g^{(2)}_{i,i}(\tau)$ becomes more complex, and therefore, we alternatively define 

\begin{equation}
    \Delta g^{(2)}_{i,i} \equiv \int d\tau \qty(g^{(2)}_{i,i}(\tau)-1).
    \label{Deltag2}
\end{equation}

The range of the temporal integral is assumed to be sufficiently large. From equation~\eqref{n_auto},

\begin{equation}
    \Delta g^{(2)}_{i,i}
    = 2\pi \int d\tau \abs{\mathcal{F}[\abs{f(\omega_i)}^2](\tau)}^2
    = 2\pi \int d\omega_i \abs{f(\omega_i)}^4.
    \label{auto_num}
\end{equation}

Substituting equation~\eqref{multi_jsa} to equation~\eqref{auto_num}, we obtain

\begin{align}
    \Delta g^{(2)}_{i,i} &= \frac{2\pi}{M^2}\sum_m \int d\omega_i \abs{f_m(\omega_i)}^4 \notag \\
    &=\frac{2\pi}{M} \qty{\int d\omega \frac{1}{\qty(\gamma_i^2+\omega^2)^2}\frac{1}{\qty(\gamma_s^2+\omega^2)^2}}
    \qty{\int d\omega \frac{1}{\gamma_i^2+\omega^2}\frac{1}{\gamma_s^2+\omega^2}}^{-2} \notag \\
    &=\frac{1}{M} \frac{\gamma_i^2+3\gamma_i\gamma_s+\gamma_s^2}{\gamma_i \gamma_s(\gamma_i+\gamma_s)} \\
    &=
        \begin{cases}
            \frac{5}{2\gamma_s M} & (\gamma_s=\gamma_i) \\
            \frac{1}{\gamma_s M} & (\gamma_s \ll \gamma_i)
        \end{cases}
    .
    \label{auto_formula}
\end{align}

Therefore, if $\gamma_i$ and $\gamma_s$ are known, the number of cavity modes can be estimated using this formula. In addition to equation~\eqref{C}, the actual value obtained from the measurement is expressed as
\begin{equation}
    C^{(2)}_{i,i} (\tau) \propto \int d\tau' g_{\sqrt{2}\sigma}(\tau-\tau') G^{(2)}_{i,i}(\tau').
    \label{C_auto}
\end{equation}

After integration over time, we obtained $\Delta g^{(2)}_{i,i}$ in equation~\eqref{Deltag2} as

\begin{align}
    \int d\tau \qty(C^{(2)}_{i,i} (\tau)-C^{(2)}_{i,i} (\tau \rightarrow \infty)) \propto \int d\tau \qty(G^{(1,1)}_{i,i}(\tau)-G^{(1,1)}_{i,i}(\tau \rightarrow \infty)) 
    \propto \Delta g^{(2)}_{i,i}.
\end{align}

Therefore, $\Delta g^{(2)}_{i, i}$ can be measured without depending on the temporal resolution of the detector.

{In our experiment, we estimate the value of $G^{(1,1)}_{i, i}(\tau \rightarrow \infty)$ from the best fit to the data far from the point of $\tau = 0$, as indicated by the orange lines in Fig.~4.
We then calculated $\Delta g^{(2)}_{i,i}$ as $2.1$~ns and $0.17$~ns, respectively. 
Combining the values of $\gamma_s$ obtained in Table.~\ref{cavity}, we estimate the number of modes $M$ to be $1.2$ and $14.9$, respectively.}

\subsection{Full data of time histogram of coincidence counts and estimated property of frequency comb}

\nolinenumbers
\begin{figure*}[tbp]
    \centering
    \begin{tabular}{cccc}
        \raisebox{-\height}{\bf a} & \raisebox{-\height}{\includegraphics[scale=0.3, trim=0 0 0 0, clip]{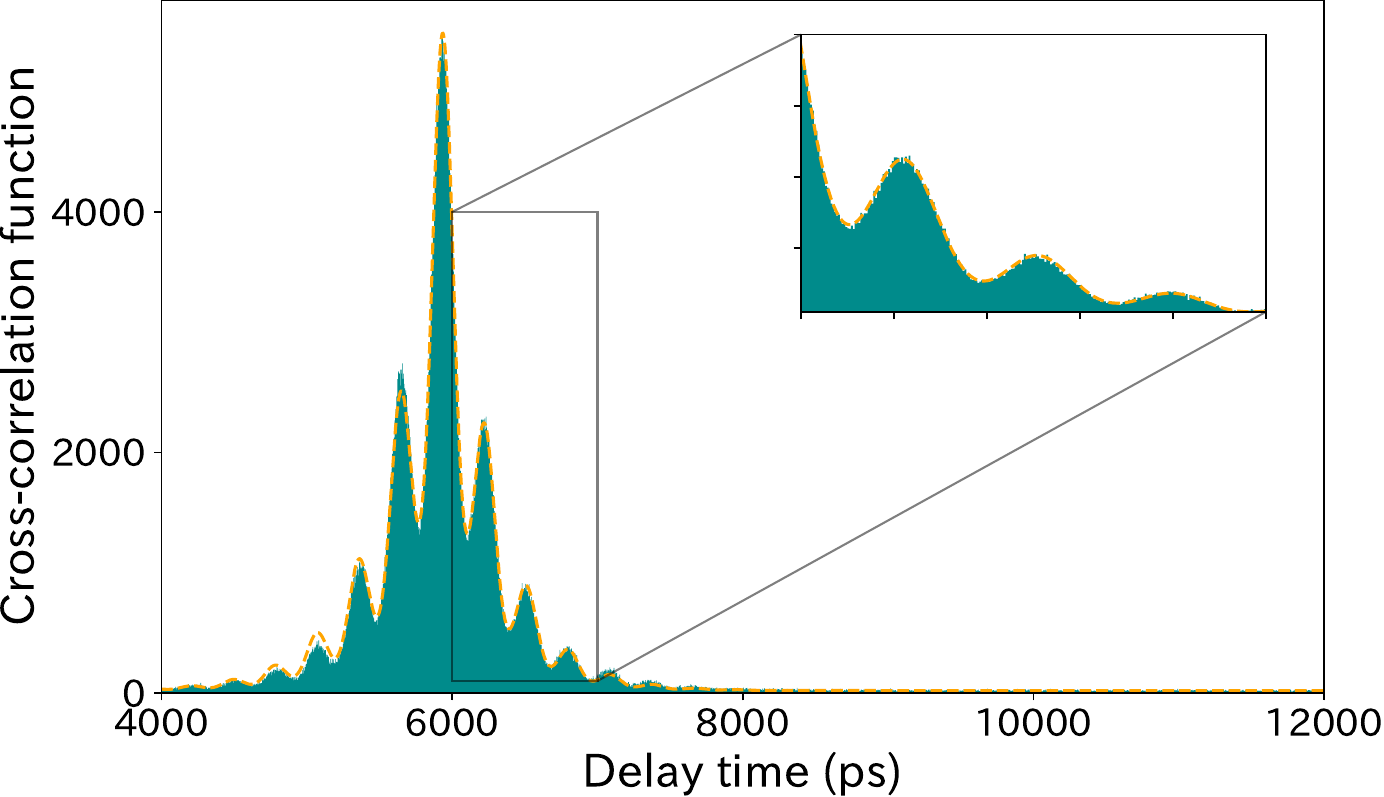}} &
        \raisebox{-\height}{\bf f} & \raisebox{-\height}{\includegraphics[scale=0.3, trim=0 0 0 0, clip]{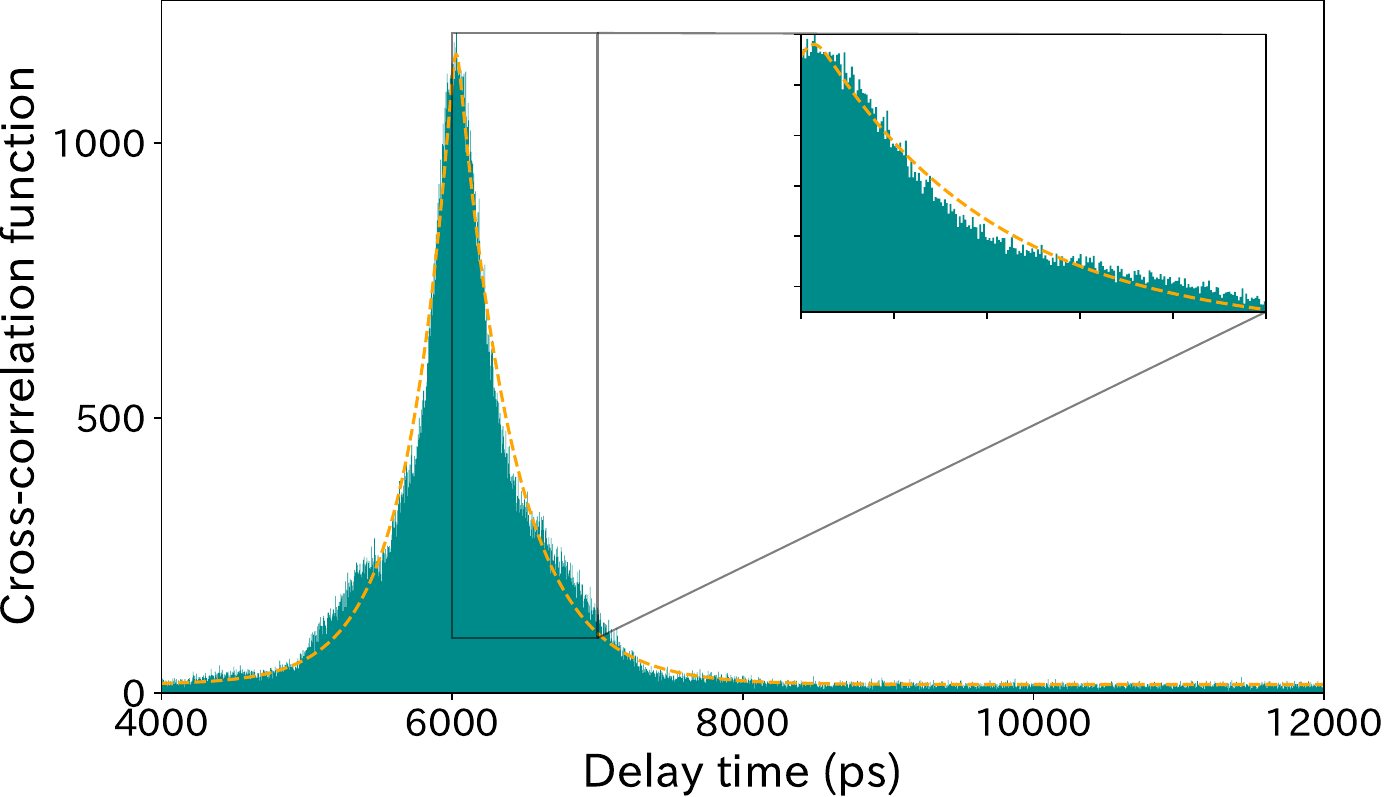}} \\
        \raisebox{-\height}{\bf b} & \raisebox{-\height}{\includegraphics[scale=0.3, trim=0 0 0 0, clip]{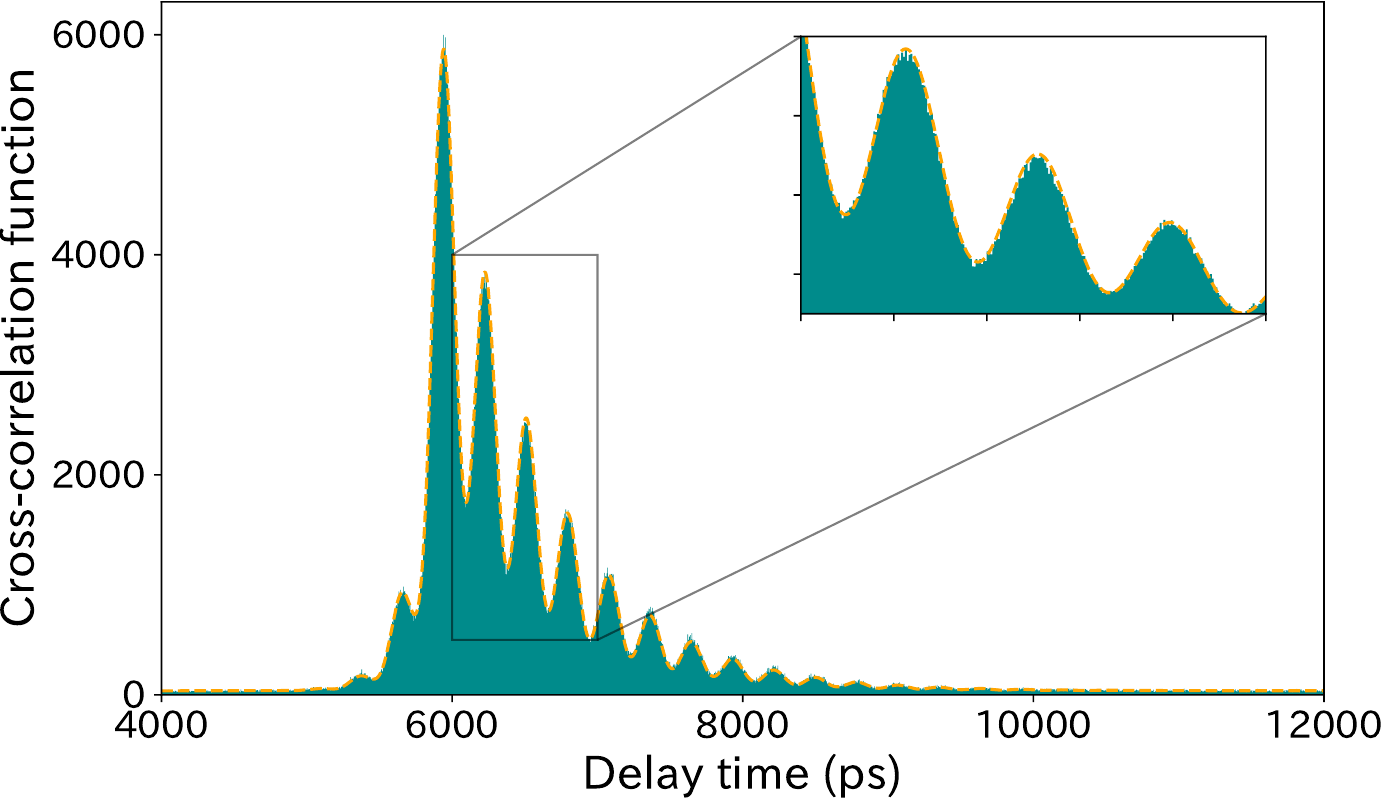}} &
        \raisebox{-\height}{\bf g} & \raisebox{-\height}{\includegraphics[scale=0.3, trim=0 0 0 0, clip]{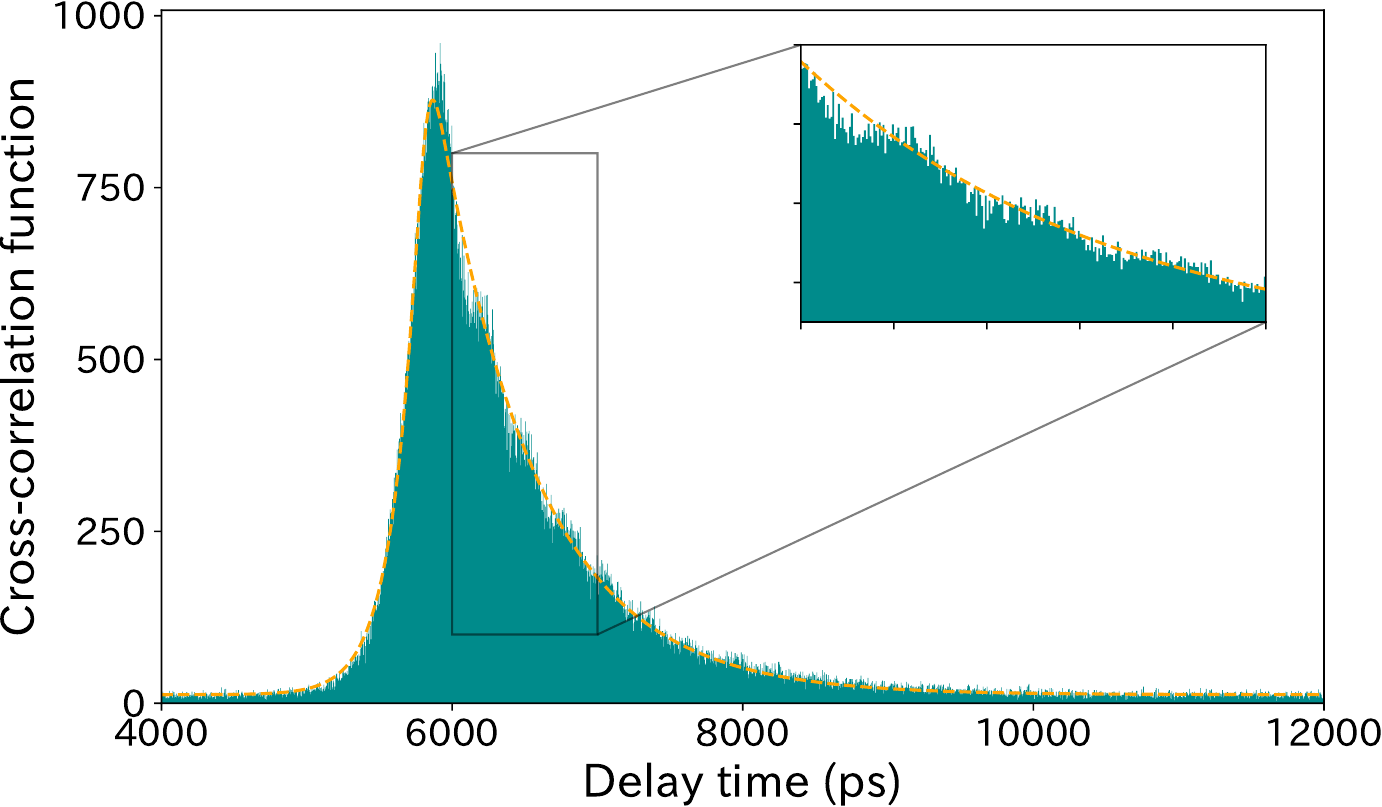}} \\
        \raisebox{-\height}{\bf c} & \raisebox{-\height}{\includegraphics[scale=0.3, trim=0 0 0 0, clip]{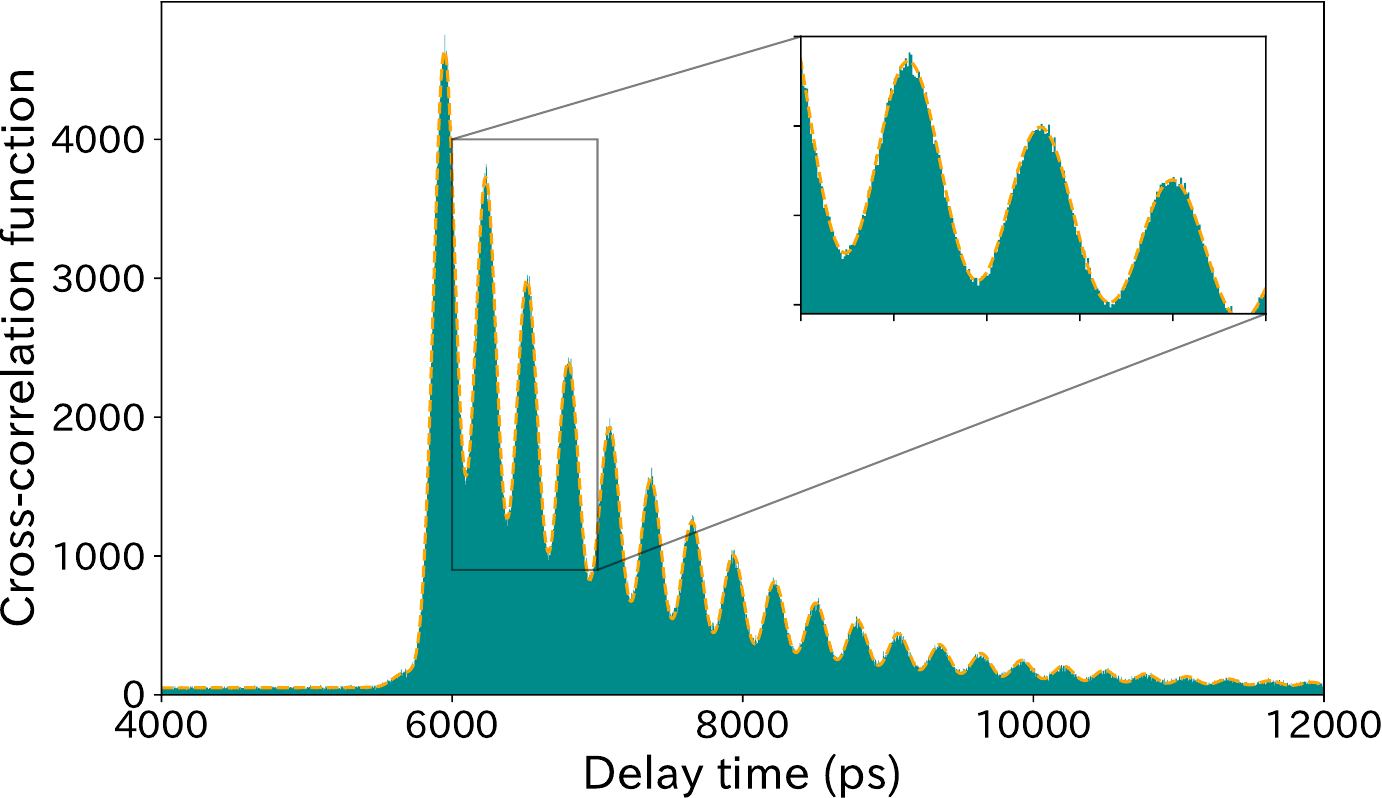}} &
        \raisebox{-\height}{\bf h} & \raisebox{-\height}{\includegraphics[scale=0.3, trim=0 0 0 0, clip]{./htm_158049154097_003_400012000-crop.pdf}} \\
        \raisebox{-\height}{\bf d} & \raisebox{-\height}{\includegraphics[scale=0.3, trim=0 0 0 0, clip]{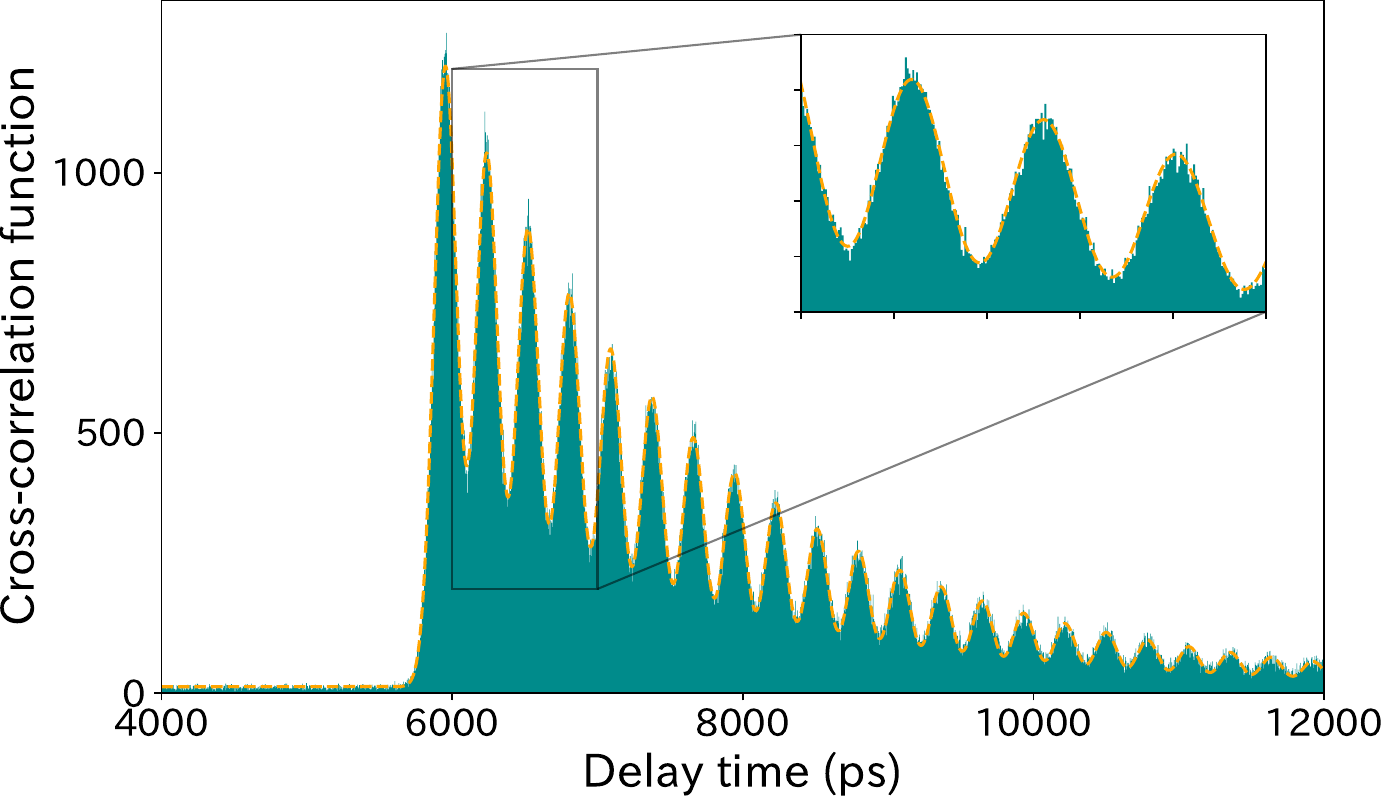}} &
        \raisebox{-\height}{\bf i} & \raisebox{-\height}{\includegraphics[scale=0.3, trim=0 0 0 0, clip]{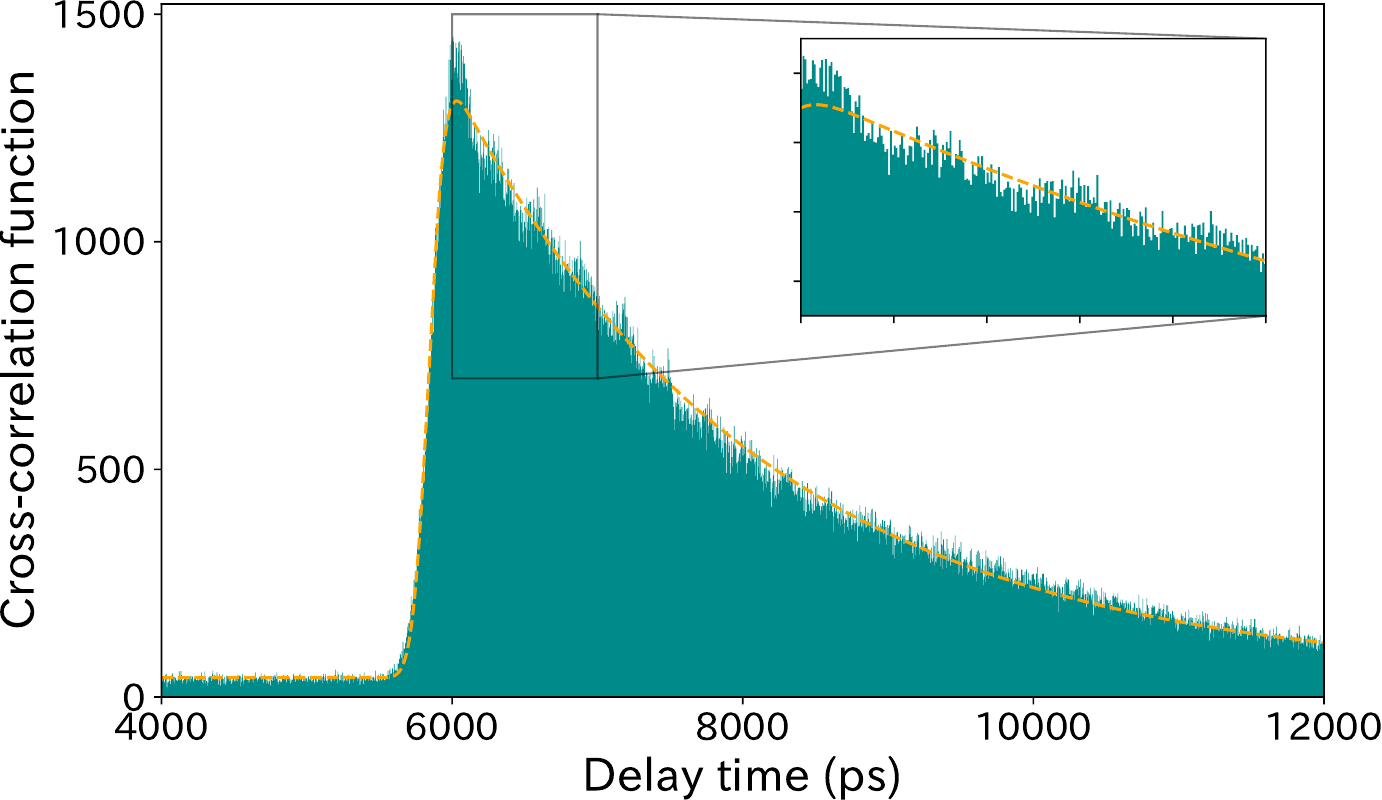}} \\
        \raisebox{-\height}{\bf e} & \raisebox{-\height}{\includegraphics[scale=0.3, trim=0 0 0 0, clip]{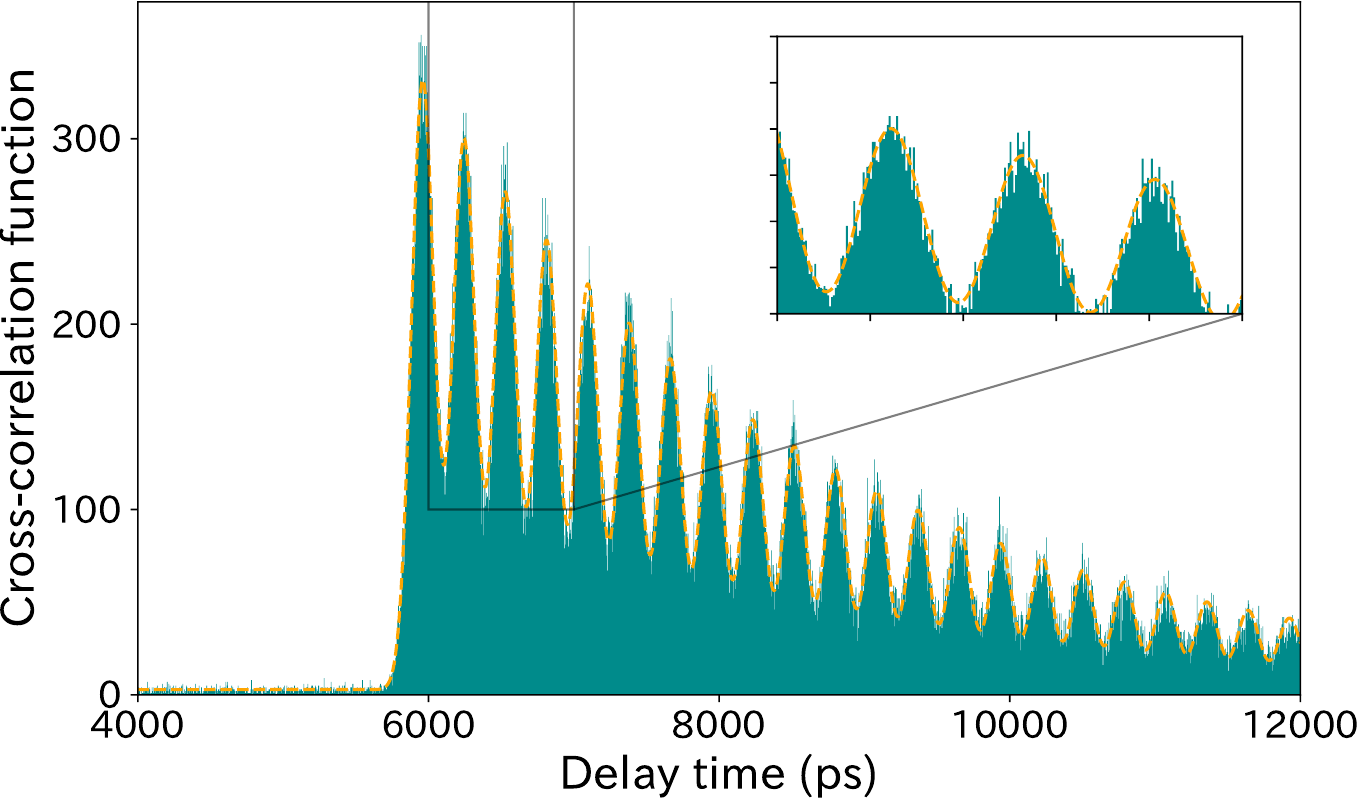}} &
        \raisebox{-\height}{\bf j} & \raisebox{-\height}{\includegraphics[scale=0.3, trim=0 0 0 0, clip]{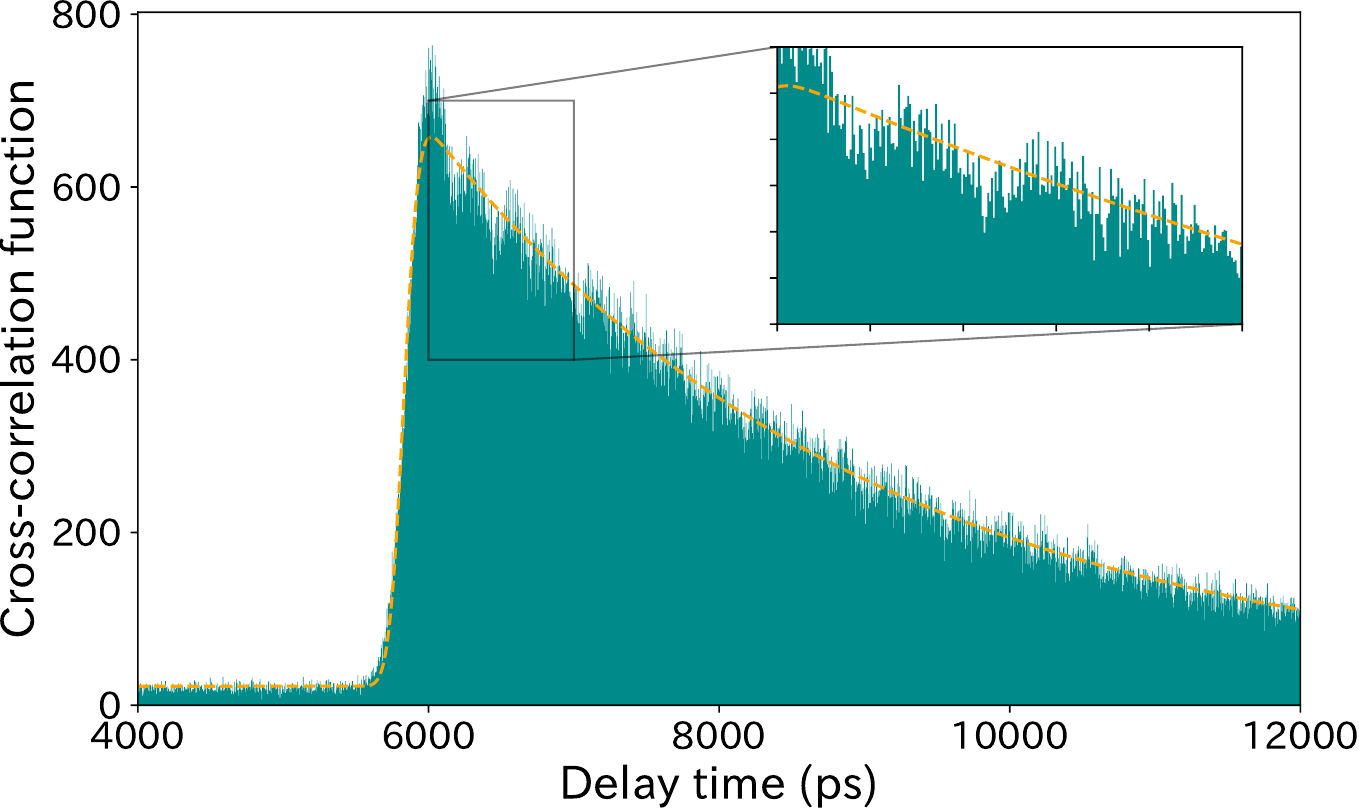}} \\
    \end{tabular}
    \caption{{\bf Full data of the temporal second-order cross-correlation.} The delayed coincidence counts between the signal and idler photons with 4-ps time resolution, which correspond to the temporal second-order cross-correlation function. Their polarization is HH, but the results for the polarization of VV are almost the same as those shown in Fig.~S3a. The bandwidths of the BPFs are {\bf a-e} 3.00~nm ($\sim$ 100 frequency mode) and {\bf f-j} 0.03~nm (single-frequency mode). The wavelengths are {\bf a, f} (1560.48~nm, 1560.48~nm), {\bf b, g} (1570.48~nm, 1550.61~nm), {\bf c, h} (1580.48~nm, 1540.98~nm), {\bf d, i} (1590.48~nm, 1531.59~nm), and {\bf e, j} (1600.48~nm, 1522.43~nm). The data recording time is 1000~s.}
\end{figure*}
We measured the time histogram of coincidence counts in each frequency range by setting the central wavelength of the BPFs to (1560.48~nm, 1560.48~nm), (1570.48~nm, 1550.61~nm), (1580.48~nm, 1540.98~nm), (1590.48~nm, 1531.59~nm), and (1600.48~nm, 1522.43~nm), and by setting the bandwidths of the BPFs to 3.00~nm and 0.03~nm. Figure~S4 shows all the results of the fitting obtained using equation~\eqref{sum} for the bandwidth of 3.00~nm and equation~\eqref{single} for the bandwidth of 0.03~nm. The fitting results indicate that $p \simeq 1$ in equation~\eqref{sum}, and the results of the fitting using equation~\eqref{sum} are almost the same as the results of the fitting using equation~\eqref{multi}. The properties of the generated frequency comb, which are estimated by the best fit to the data, are summarized in Table S1. The FWHM depends on the frequency because it is related to the frequency-dependent reflectance of PPLN/WR~\cite{Ikuta2019d}.

\nolinenumbers
\begin{table}[tbp]
    \centering
    \caption{{\bf Estimated properties of the generated frequency comb.} The FSRs and FWHMs of the generated frequency combs in different frequency ranges are estimated from the best fit to the temporal cross-correlation measurements. Finesse (FSR/FWHM) and Q factor (frequency/FWHM) are calculated from these values. () indicates the data of the idler photon in the doubly resonant case.}
    \begin{tabular}{|r||r|r|r|r|r|r|}
        \hline
        \multicolumn{1}{|c||}{Wavelength} & \multicolumn{1}{c|}{FSR} &
        \multicolumn{1}{c|}{FWHM} & \multicolumn{1}{c|}{Finesse} &
        \multicolumn{1}{c|}{Q factor} \\ \hline \hline
        1560~nm & 3.5~GHz & 516(453)~MHz & 7(8) & 3.7(4.2) $\times 10^5$ \\ \hline
        1570~nm & 3.5~GHz & 243(1043)~MHz & 15(3) & 7.9(1.8) $\times 10^5$ \\ \hline
        1580~nm & 3.5~GHz & 126~MHz & 28 & 1.5 $\times 10^6$ \\ \hline
        1590~nm & 3.5~GHz & 85~MHz & 41 & 2.2 $\times 10^6$ \\ \hline
        1600~nm & 3.5~GHz & 57~MHz & 62 & 3.3 $\times 10^6$ \\ \hline
    \end{tabular}
    \label{cavity}
\end{table}

\begin{figure}[tbp]
    \centering
    \includegraphics[scale=0.7, trim=0 0 0 0, clip]{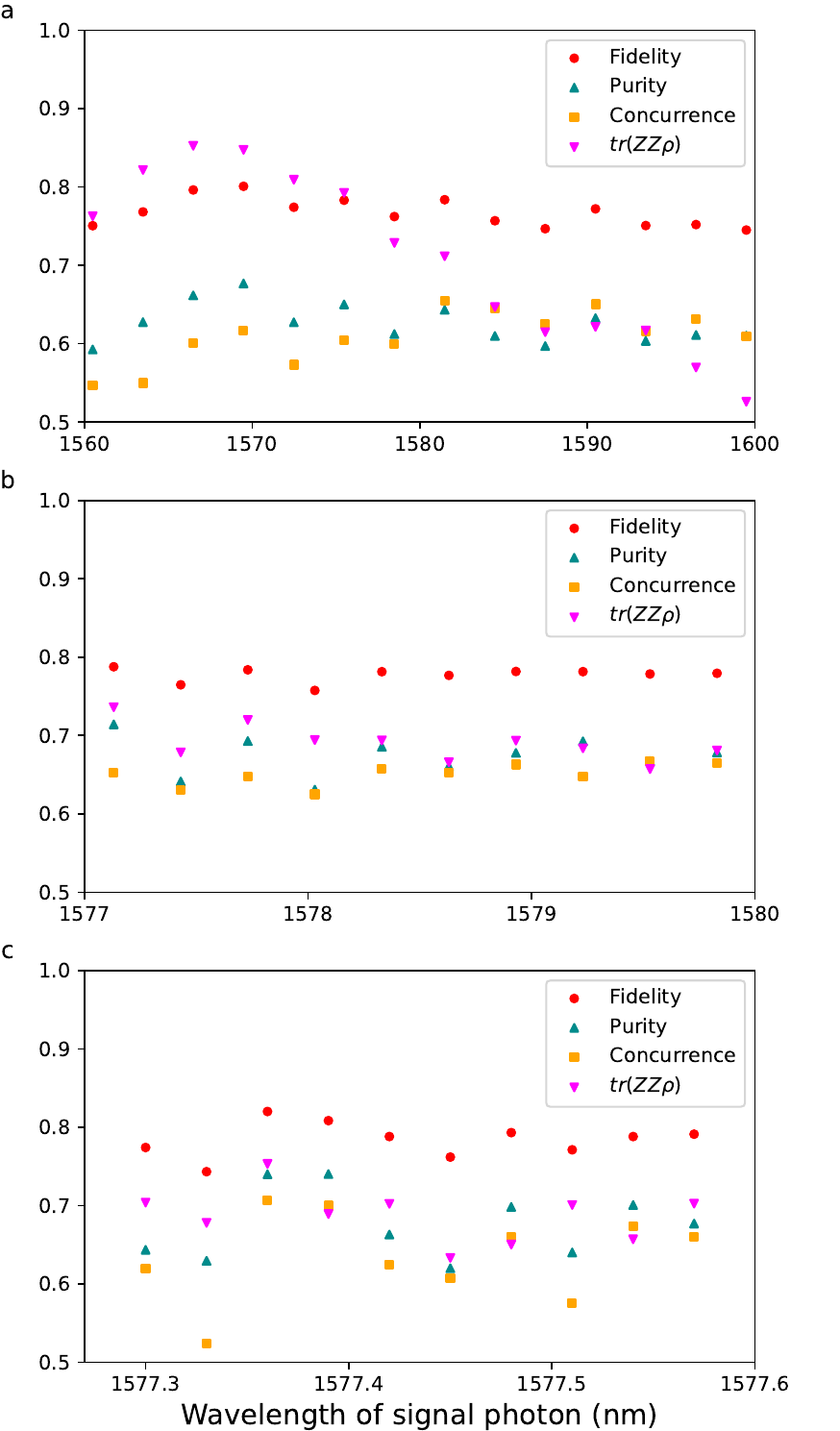}
    \caption{{{\bf Characterizations of the generated entangled states.} The fidelity, purity, concurrence, and $tr(ZZ\rho)$ are calculated from the reconstructed density matrices of the polarization-entangled states. The frequency ranges of a--c are the same as those shown in Fig.~5a-c in the manuscript.}}
    \label{entanglement_measure}
\end{figure}
\subsection{Another method for generating the polarization entangled state using all polarization components}
From equation~\eqref{4output}, we can utilize the entire state as the polarization-entangled state by setting the length of the arms of the Sagnac interferometer to the same values: $\tau_r=\tau_l$ with the stabilization of the relative phase. The inner product of the two terms in equation~\eqref{4output} satisfies

\begin{equation}
    \qty(\bra{V}_{s}+\beta_H^* e^{i\Delta \theta}\bra{H}_{s})\qty(\ket{H}_{s}+\beta_V e^{i\Delta \theta}\ket{V}_{s}) 
    \propto r^*_l t_r \qty(\eta_{sl}^2-\eta_{sr}^2),
\end{equation}

where we used the relation $r^*_l t_r + t^*_l r_r = 0$.
Therefore, in the symmetric loss case, that is, $\eta_{sl} = \eta_{sr}$, the state in equation~\eqref{4output} represents a maximally entangled state. However, asymmetric losses degrade orthogonality and cause a decrease in the fidelity to the maximally entangled state. The lower bound of its fidelity is

\begin{align}
    F &\geq \frac{1}{2} \qty[1+ \abs{1-\beta_H \beta_V}\middle/ {\sqrt{\qty(1+\abs{\beta_H}^2)\qty(1+\abs{\beta_V}^2)}}], \notag \\
    &= \frac{1}{2}\qty[1+ \qty{\qty(1+\abs{r}^2\qty(\frac{\eta_{sl}^2}{\eta_{sr}^2}-1))\qty(1+\abs{r}^2\qty(\frac{\eta_{sr}^2}{\eta_{sl}^2}-1))}^{-1/2}].
\end{align}

In the second equation, $\abs{r_r}=\abs{r_l}\equiv\abs{r}, \ \abs{t_r}=\abs{t_l}\equiv\abs{t}, \ \abs{r}^2+\abs{t}^2=1$, and $1-\beta_H \beta_V =1 - r_r r_l / t_r t_l = \abs{t}^{-2}$. This lower bound is almost 1 in the typical experiments, and this method works well. This approach has twice the generation rate and does not require a temporal distinction.


\section{References}
\bibliography{sagnac.bib}